\documentclass[12pt]{article}

\usepackage{jheppub}
\usepackage{tabularray}
\usepackage{slashed}
\usepackage{float}
\usepackage{cleveref}

\graphicspath{{./figures/}}

\newcommand{\centeredgraphics}[2][]{\vcenter{\hbox{\includegraphics[#1]{#2}}}}

\def\als{\alpha_{\rm s}}
\def\al{\alpha}
\def\dsl{\,\raise.15ex\hbox{/}\mkern-13.5mu D}

\def\MS{\overline{\rm MS}}
\def\one{{(1)}}
\def\two{{(2)}}
\def\vp{{\bf p}}
\def\vk{{\bf k}}

\def\lla{\langle\!\langle}
\def\rra{\rangle\!\rangle}

\def\bfsigma{\mbox{\boldmath $\sigma$}}
\def\bfnabla{\mbox{\boldmath $\nabla$}}
\def\bfgamma{\mbox{\boldmath $\gamma$}}
\newcommand{\bfm}[1]{\mbox{\boldmath$#1$}}

\newcommand{\nn}{\nonumber}

\begin{document}
\title{\boldmath The ultrafine splitting of heavy quarkonium with 
next-to-next-to-next-to-next-to-leading-\\order accuracy}

\author{Jose M. Escario$^{a}$,} \author{Andreas Maier$^{b}$,} \author{Clara Peset$^{a}$,} \author{Antonio Pineda$^{c,b}$} 
\affiliation{ ${}^a$ Departamento de F\'{\i}sica Te\'orica \& IPARCOS, Universidad Complutense de Madrid,\\ Plaza de las Ciencias 1, 28040 Madrid, Spain}
\affiliation{${}^b$ Institut de Física d'Altes Energies (IFAE),  The Barcelona Institute of Science and \\Technology, Campus UAB, E-08193 Bellaterra (Barcelona), Spain}
\affiliation{ ${}^c$ Departament de Física, Universitat Autònoma de Barcelona, E-08193 Bellaterra, Spain}

\emailAdd{joescari@ucm.es}
\emailAdd{amaier@ifae.es}
\emailAdd{cpeset@ucm.es}
\emailAdd{AntonioMiguel.Pineda@uab.es}
\abstract{We compute the hyperfine splitting of P-wave heavy quarkonium states with next-to-next-to-next-to-next-to-leading-order accuracy. The resummation of logarithms with next-to-next-to-next-to-next-to-leading-logarithmic accuracy is also addressed. A phenomenological analysis of these results is performed for bottomonium, charmonium and the $B_c$ system. We also apply these results to positronium, muonium, hydrogen and muonic hydrogen.}

\preprint{IPARCOS-UCM-26-012}

\maketitle




\section{Introduction}

The combined use of non-relativistic effective field theories \cite{Caswell:1985ui,Bodwin:1994jh,Pineda:1997bj,Brambilla:1999xf} (for some reviews see \cite{Brambilla:2004jw,Pineda:2011dg}) and efficient multiloop computations has made it possible to achieve very high precision in the computation of heavy quarkonium spectroscopy \cite{Kniehl:1999ud,Brambilla:1999xj,Pineda:2001ra,Kniehl:2002br,Kniehl:2003ap,Penin:2004xi,Beneke:2005hg,Kiyo:2014uca,Peset:2015vvi,Peset:2018jkf}.

The aim of this paper is to establish the foundations for achieving next-to-next-to-next-to-next-to-leading-order (N$^4$LO) accuracy and next-to-next-to-next-to-next-to-leading-logarithmic accuracy (N$^4$LL) in spin-dependent observables (in particular, energy splittings) of heavy quarkonium. 

In order to reach this level of precision, the following results and computations are required:
\begin{enumerate}
\item
The Wilson coefficients of the dimension six four-fermion operators at two-loops. At one loop, they were computed in \cite{Pineda:1998kj}. Their expression in $D$-dimensions can be found in Ref.~\cite{Beneke:2013jia}. The two-loop result in $D$ dimensions was obtained in \cite{Gerlach:2019kfo}.
\item
The leading renormalization group running of the spin-dependent and velocity-dependent dimension-eight four-fermion operators. See, for example, \cite{Bodwin:1994jh}.
\item
The chromomagnetic moment $c_F$ to two loops, as well as its renormalization group running. This quantity has been computed up to order $\alpha_s^3$ \cite{Amoros:1997rx,Czarnecki:1997dz,Grozin:2007fh}, together with its anomalous dimension.
\item
The $1/m^2$ spin-dependent potential to two-loops in $D$-dimensions. In this paper, we perform such a computation for the spin-dependent and velocity-independent potential and compare it with the existing results in \cite{Penin:2004xi,Mishima:2024afk}.
\item
The $1/m^3$ spin-dependent potential to one loop in $D$-dimensions. In this paper, we perform such computation for the spin-dependent and velocity-independent potential and compare with existing results in \cite{Penin:2004xi,Mishima:2024afk}.
\item
The spin-dependent $1/m^4$ potential at tree-level. Here, we include additional potentials beyond those considered in \cite{Penin:2004xi}, which are relevant for P-wave observables or contribute to N$^4$LO.
\item
Quantum-mechanical perturbation theory applied to the resulting potential to the required level of accuracy. 
\item
Ultrasoft effects must also be considered. Even if they do not contribute to the hyperfine splitting to ${\cal O}(m\alpha_s^6)$, logarithmically enhanced effects appear at ${\cal O}(m\alpha_s^7)$. For instance, they contribute to S-wave hyperfine splitting at N$^3$LL and to P-wave hyperfine splitting at N$^4$LL.
\item
A prescription for handling Dirac and Pauli matrices in $D$-dimensions in a consistent way across the different computations above. This issue is nontrivial, as these structures arise from different energy scales.
\item 
A comprehensive renormalization group analysis of the previous computations. 
\end{enumerate}

In this paper, we address several of the points listed above. We then apply the resulting findings to the hyperfine splitting of P-wave states. Other observables will be considered in forthcoming papers. 

We define the P-wave hyperfine splitting in terms of the center of gravity (c.o.g.), namely as
\begin{align}
     \Delta \equiv E(^{1}P_1)-E(^{3}P)_{c.o.g}=E(^{1}P_1)-\frac{1}{9}\left(5E(^{3}P_2)+3E(^{3}P_1)+E(^{3}P_0)\right).
\end{align}
Its experimental values read \cite{ParticleDataGroup:2024cfk} (as given in Ref.\ \cite{Peset:2018jkf})
\begin{align}
     \Delta^{b\bar b,1P}_{\rm exp}=-0.57(84) \, {\rm MeV}\;, \quad  \Delta^{c\bar c,1P}_{\rm exp}=+0.08(13) \, {\rm MeV}\;, \quad  \Delta^{b\bar b,2P}_{\rm exp}=-0.44(1.26) \, {\rm MeV}\,,
\end{align}
where the superscripts $n_r P$ denote states with angular momentum $l=1$ and principal quantum numbers $n = n_r + 1$. 
The experimental value for $B_c$ is at present unknown. 

$\Delta$ is often referred to as the ultrafine splitting in the literature. The reason lies in its small value. This energy shift starts at ${\cal O}(m\als^5)$. It was computed up to N$^3$LO for equal masses in \cite{Titard:1994id} and with N$^3$LL accuracy in Ref.\ \cite{Peset:2018jkf} for equal or unequal masses. At this order, the quantity is positive, as it is one of the few places where light-fermion effects dominate over non-Abelian contributions. Another aspect that makes $\Delta$ particularly interesting is its proportionality to $\als^5$. This makes it a potential candidate for precise determinations of $\als$, provided the perturbative series shows good convergence and experimental precision improves. In this respect, the $1P$ bottomonium splitting is especially promising. In this paper, we  compute $\Delta$ with N$^4$LO accuracy and partially incorporate the hard logarithms entering the N$^4$LL evaluation. Both are novel results. 

Since this paper focuses on the P-wave hyperfine splitting, pure hard effects coming from the dimension-six local four-fermion operators in NRQCD do not contribute. On the other hand, hard effects arising from the Wilson coefficients of the bilinear terms of the NRQCD Lagrangian do contribute, as do those from dimension-eight local four-fermion operators. The former are included in the present analysis, whereas the latter will be incorporated in a forthcoming paper. Ultrasoft effects are also not included here. These contribute to S-wave hyperfine splitting at N$^3$LL and to P-wave hyperfine splitting at N$^4$LL, and will likewise be addressed in a forthcoming paper.

$\Delta$ is also of interest in the context of precision studies in atomic physics. For equal mass states such as positronium, the first nonvanishing contributions appear at ${\cal O}(m\al^6)$, already at N$^4$LO, which has motivated searches for new physics \cite{Lamm:2017lrn,Jentschura:2020zlr}. The quantity $\Delta$ has been computed in Refs.~\cite{Khriplovich:1993zz,Czarnecki:1999mw,Zatorski} up to ${\cal O}(m\al^6)$ for the case of positronium. This result was recently challenged in Ref.\ \cite{Patkos:2024lqf} (the unequal mass case has also been discussed there). The current experimental precision \cite{Gurung:2020hms,Hagena,Ley}:
 \begin{align}
     \Delta^{pos}_{\rm exp}=-4.5(9.1) \, {\rm MHz}
 \,
 \end{align}
 does not match the theoretical accuracy. Therefore, the experimental result, whose dominant uncertainty comes from the 2 $^{1}P_1$ state \cite{Ley}, cannot yet resolve this discrepancy. 

By taking the Abelian limit, our results can be applied to QED systems for point-like particles such as positronium/dimuonium or muonium. For these, our computation disagrees with the result obtained in Ref.\ \cite{Patkos:2024lqf} but agrees with the earlier results. After communicating our findings to the authors of Ref.\ \cite{Patkos:2024lqf}, a mistake in their calculation was identified, and their updated results are now in agreement with the results in Refs.~\cite{Khriplovich:1993zz,Czarnecki:1999mw,Zatorski}.

Our results can also be directly applied to hydrogen-like system,\footnote{Even though for systems with very different masses, $\Delta$ is not commonly used as an observable.} neglecting hadronic effects, and to muonic hydrogen in the limit of small electron mass.

The structure of the paper goes as follows. In Sec.~\ref{Sec:pNRQCD} we present the pNRQCD Lagrangian. In Sec.~\ref{Sec:Pot}, we derive the potentials relevant for this work.  We first compute the $1/m^2$ spin-dependent and velocity-independent potential to two-loops in the Wilson-loop matching scheme for general $D$-dimensions. We then compute the $1/m^3$ spin-dependent potential to one loop for general $D$-dimensions in the off-shell matching scheme, as well as the $1/m^4$ tree-level potential. We also discuss the relation between the different matching schemes in this section. Using these results, in Sec.~\ref{Sec:Delta}, we compute $\Delta$ with N$^4$LO precision and also provide a partial N$^4$LL result that incorporates hard logarithms. The final results are given in Sec. \ref{Sec:FunalResult}, together with a brief phenomenological analysis. In Sec.~\ref{sec:QED} we apply our results to QED and the final section is devoted to the conclusions. Additional details are provided in the appendices. In Appendix~\ref{App:Spin}, we describe our treatment of Dirac and Pauli matrix traces in $D$-dimensions. For completeness, some Feynman rules are collected in Appendix~\ref{App:FR}. In Appendix~\ref{App:ME} we present expectation values of the potentials required for the computation of $\Delta$.

\section{pNRQCD Lagrangian}
\label{Sec:pNRQCD}


Integrating out the soft modes in NRQCD, we end up with the EFT named pNRQCD.
The most general pNRQCD Lagrangian 
compatible with the symmetries of QCD that can be constructed
with a singlet and an octet (quarkonium) field, as well as an ultrasoft gluon field to NLO in the 
multipole expansion, has the form~\cite{Pineda:1997bj,Brambilla:1999xf}
\begin{align}
    &  \!\!\!\!\!
{\cal L}_{\rm pNRQCD} = \!\! \int \!\! d^3{\bf r} \; {\rm Tr} \,  
\Biggl\{ {\rm S}^\dagger \left( i\partial_0 
- h_s({\bf r}, {\bf p}, {\bf P}_{\bf R}, {\bf S}_1,{\bf S}_2) \right) {\rm S} 
+ {\rm O}^\dagger \left( iD_0 
- h_o({\bf r}, {\bf p}, {\bf P}_{\bf R}, {\bf S}_1,{\bf S}_2) \right) {\rm O} \Biggr\}
\nn
\\
& \qquad\qquad 
+ V_A ( r) {\rm Tr} \left\{  {\rm O}^\dagger {\bf r} \cdot g{\bf E} \,{\rm S}
+ {\rm S}^\dagger {\bf r} \cdot g{\bf E} \,{\rm O} \right\} 
+ \frac{V_B (r)}{ 2} {\rm Tr} \left\{  {\rm O}^\dagger {\bf r} \cdot g{\bf E} \, {\rm O} 
+ {\rm O}^\dagger {\rm O} {\bf r} \cdot g{\bf E}  \right\}  
\nn
\\
& \qquad\qquad 
- \frac{1}{ 4} G_{\mu \nu}^{a} G^{\mu \nu \, a} 
+  \sum_{i=1}^{n_l} \bar q_i \, i \dsl \, q_i 
\,,
\label{Lpnrqcd}
\\
& 
\nn 
\\
& 
h_s({\bf r}, {\bf p}, {\bf P}_{\bf R}, {\bf S}_1,{\bf S}_2) = 
 \frac{{\bf p}^2 }{ 2\, m_{ r}}
+ 
\frac{{\bf P}_{\bf R}^2 }{ 2\, M} + 
V_s({\bf r}, {\bf p}, {\bf P}_{\bf R}, {\bf S}_1,{\bf S}_2), 
\\
& 
h_o({\bf r}, {\bf p}, {\bf P}_{\bf R}, {\bf S}_1,{\bf S}_2) = 
\frac{{\bf p}^2 }{ 2\, m_{ r}}
+ 
\frac{{\bf P}_{\bf R}^2 }{ 2\,M}  + 
V_o({\bf r}, {\bf p}, {\bf P}_{\bf R}, {\bf S}_1,{\bf S}_2), 
\end{align}
where $iD_0 {\rm O} \equiv i \partial_0 {\rm O} - g [A_0({\bf R},t),{\rm O}]$, 
${\bf P}_{\bf R} = -i{\bfnabla}_{\bf R}$ for the singlet,  
${\bf P}_{\bf R} = -i{\bf D}_{\bf R}$ for the octet (where the covariant derivative is in the adjoint representation), 
${\bf p} = -i\bfnabla_{\bf r}$,
\begin{align}
m_{r} = \frac{m_1 m_2}{m_1+m_2}
\end{align}
and $M = m_1+m_2$. 
We adopt the color normalization  
\begin{align}
    {\rm S} = { S\, 1\!\!{\rm l}_c / \sqrt{N_c}} \,, \quad\quad\quad 
{\rm O} = O^a { {\rm T}^a / \sqrt{T_F}}\,,
\label{SSOO}
\end{align}
for the singlet field $S({\bf r}, {\bf R}, t)$ and the octet field $O^a({\bf r}, {\bf R}, t)$.
Here and throughout this paper we denote the quark-antiquark distance vector by ${\bf r}$, the center-of-mass position of the quark-antiquark system by ${\bf R}$, and the time by $t$.

Both $h_s$ and the potential $V_s$ are operators acting on the Hilbert space of a heavy quark-antiquark system in the singlet configuration.\footnote{Therefore, in a more mathematical notation: $h \rightarrow \hat h$, $V_s({\bf r},{\bf p}) \rightarrow \hat V_s(\hat {\bf r},\hat {\bf p})$. We will however avoid this notation in order to facilitate the reading.}
$V_s$ (and $V_o$) can be Taylor expanded in powers of $1/m$ (up to logarithms). At low orders we have
\begin{align}
     V_s &=
V^{(0)} +\frac{V^{(1)}}{m_r}+ \frac{V_{{\bf L}^2}^{(2)}}{m_1m_2}\frac{{\bf L}^2 }{ r^2}
+\frac{1}{ 2m_1m_2}
\left\{{\bf p}^2,V^{(2)}_{{\bf p}^2}(r)\right\}+\frac{V_r^{(2)}}{m_1m_2}
\nn
\\
&+\frac{1}{m_1m_2} V_{S^2}^{(1,1)}(r){\bf S}_1\cdot{\bf S}_2+\frac{1}{m_1m_2} V_{{\bf S}_{12}}^{(1,1)}(r){\bf
  S}_{12}({\bf r})
  \nn
  \\
  &
  +\frac{1}{m_1m_2}V^{(2)}_{LS_1}(r){\bf L}\cdot{\bf S}_1 +\frac{1}{m_1m_2}V^{(2)}_{LS_2}(r) {\bf L}\cdot{\bf S}_2
+{\cal O}(1/m^3),
\label{V1ovm2}
\end{align}
where, ${\bf S}_1=\bfsigma_1/2$, ${\bf S}_2=\bfsigma_2/2$, ${\bf L} \equiv {\bf r} \times {\bf p}$, and $\displaystyle{{\bf S}_{12}({\bf r}) \equiv \frac{
3 { {\bf r}}\cdot \bfsigma_1 \,{ {\bf r}}\cdot \bfsigma_2}{r^2} - \bfsigma_1\cdot \bfsigma_2}$.

The above potentials can also be written as follows
\begin{align}
    V^{(1)}=V^{(1,0)}(r)=V^{(0,1)}\equiv - \frac{C_FC_A D^{(1)}}{4 r^2}
\,,
\end{align}

\begin{align}\label{delVL2}
    \frac{V_{{\bf L}^2}^{(2)}}{m_1m_2}\equiv \frac{V_{{\bf L}^2}^{(2,0)}(r)}{m_1^2}+\frac{V_{{\bf L}^2}^{(0,2)}(r)}{m_2^2}+\frac{V_{{\bf L}^2}^{(1,1)}(r)}{m_1m_2}\equiv \frac{C_F D_2^{(2)}}{2 m_1 m_2 r}
\,,
\end{align}
\begin{align}\label{delVp2}
    \frac{V_{{\bf p}^2}^{(2)}}{m_1m_2}\equiv \frac{V_{{\bf p}^2}^{(2,0)}(r)}{m_1^2}+\frac{V_{{\bf p}^2}^{(0,2)}(r)}{m_2^2}+\frac{V_{{\bf p}^2}^{(1,1)}(r)}{m_1m_2}\equiv -\frac{C_F D_1^{(2)}}{ m_1 m_2 r}
\,.
\end{align}
\begin{align}\label{delVLS1}
    \frac{1}{m_1m_2}V^{(2)}_{LS_1}(r)
&\equiv 
\left(\frac{1}{m_1^2}V^{(2,0)}_{LS}(r)+\frac{1}{m_1m_2}V_{L_2S_1}^{(1,1)}(r)
\right) \equiv \frac{3C_F D_{LS_1}^{(2)}}{2m_1m_2}
\,,
\\
\frac{1}{m_1m_2}V^{(2)}_{LS_2}(r)\label{delVLS2}
&\equiv 
\left(\frac{1}{m_2^2}V^{(0,2)}_{LS}(r)+\frac{1}{m_1m_2}
V_{L_1S_2}^{(1,1)}(r) \right) \equiv \frac{3C_F D_{LS_2}^{(2)}}{2m_1m_2}
\,,
\end{align}
making explicit the complete power-like mass dependence. The middle definitions in \cref{delVL2,delVp2,delVLS1,delVLS2} are relevant for the unequal mass case as treated, e.g., in~\cite{Peset:2015vvi}, while the first and last definitions have been used for both equal and unequal mass cases in, e.g., ~\cite{Anzai:2018eua}.

The spin and angular-momentum dependence in the expressions above are written for $D=4$. For divergent loops, however, the potentials must be considered in dimensions different from four, in which case their interpretation becomes ambiguous. The treatment of spin is described in Appendix \ref{App:Spin}, while angular momentum is handled using standard dimensional regularization techniques in momentum space.   

The explicit expressions for the above potentials at one loop can be found in Ref.~\cite{Peset:2015vvi} for different matching schemes. We will only display them explicitly below when needed. Higher-order potentials in the $1/m$ expansion will be discussed in the following section, where relevant.

\section{Potentials}
\label{Sec:Pot}

For the purposes of this paper, we neglect ultrasoft effects and, consequently, the octet field. The problem then reduces to solving a Schrödinger equation, whose Hamiltonian can be summarized by the following expression:
\begin{align}
     h_s = -\frac{\bfnabla^2}{2m_r}+V_s =-\frac{\bfnabla^2}{2m_r}-\frac{C_F\als}{r}+\delta h.
\end{align}
$\delta h$ encodes the terms that will be relevant for the present computation. We will mainly work with the Fourier transform of $\delta h$, i.e. with $ \delta \tilde h$. In the following, we discuss potentials relevant for spin-dependent S-wave or P-wave (or, more generically, any $l \not =0$) computations. We provide some of their expressions in $D$-dimensions, as they are needed for the determination the S-wave hyperfine splitting with N$^4$LO accuracy. For the purposes of the present paper, the four-dimensional expressions are sufficient, since the result of quantum-mechanical perturbation theory is finite.

\subsection{Energy-dependent terms and associated matching-scheme dependence}

The standard form of the potential that appears in the pNRQCD Lagrangian is energy-independent. In the off-shell matching scheme, energy-dependent potentials may appear in intermediate computations. In the calculation of the spin-dependent potential made in Ref.~\cite{Penin:2004ay}, an energy-dependent $1/m^2$ potential was generated at one-loop:
\begin{eqnarray}
\delta \tilde V^{(2)}_{S^2,E}&=&
+\frac{1}{ 24}(\pi\alpha_s)^2 C_F C_A
\frac{c_F^{(1)}}{ m_1}\frac{c_F^{(2)}}{ m_2} \frac{1}{ |{\bfm k}|}
\left(-2E
\right)\bfm{\sigma}_1\cdot \bfm{\sigma}_2
\,,
\label{VsoftE}
\end{eqnarray}
where $E$ is the two-particle energy and $c_F^{(i)}, i\in\{1,2\}$, is the chromomagnetic moment of the particle with mass $m_i$.  It can be found in \cref{defcF} with NLL accuracy, which is enough for the purpose of this work. 

We have recomputed this quantity in $D=d+1$ dimensions. Its contribution to the hyperfine splitting reads: 
\begin{align}
    \langle  \delta \tilde V^{(2)}_{S^2,E}  \rangle
_\text{HF}
=
    &
-
\*\frac{g_B^4}{16} {|\bf k|}^{d-3}E
\left\{
- \frac{4}{d}\*C_F\*C_A\*\frac{c_F^{(1)}}{m_1}\*\frac{c_F^{(2)}}{m_2}\*\frac{1}{\lvert \mathbf{k} \rvert} \*(d-1)
\*i\*I_d\right\}
\,,
\label{V1loopdE}
\end{align}
  where
\begin{equation}
\label{Id}
  I_d = \frac{i}{(4\*\pi)^{\frac{d}{2}}}\*\frac{\Gamma\left(\frac{d}{2}-1\right)^2\*\Gamma\left(2-\frac{d}{2}\right)}{\Gamma(d-2)}\underset{d\to 3}{=}  \frac{i}{8} 
  \,.
\end{equation}
In \cref{V1loopdE} and throughout this paper, we define the spin-projected expectation values 
\begin{equation}
\langle V \rangle_\text{HF} \equiv \langle nl | V |nl \rangle \bigr|_{s=0} -
\langle nl | V |nl \rangle \bigr|_{s=1},\quad \langle V \rangle_\text{UF} \equiv \langle nP | V |nP \rangle \bigr|_{s=0} -
\langle nP | V |nP \rangle \bigr|_{s=1},
\end{equation}
where $|nl \rangle$ denotes the l-wave state with principal quantum
number $n$ and spin $s$.  $\langle V \rangle_\text{HF}$ denotes the projection of the potential $V$ onto the hyperfine splitting (independently of the orbital quantum number), and $\langle V \rangle_\text{UF}$ denotes the projection of the potential onto the ultrafine splitting specifically.

The energy-dependent term in \cref{V1loopdE} can be eliminated using field redefinitions, which, at leading order, is equivalent to applying the equations of motion (in our case, these include the Coulomb potential). Doing so yields two terms: 
\begin{align}
   \langle  \delta \tilde V^{(2,a)}_{S^2,E}  \rangle_\text{HF}
=
    &
-
\*\frac{g_B^4}{16} {|\bf k|}^{d-3}
\frac{\mathbf{p}^2+\mathbf{p'}^2}{2\*m_r}
\left\{
- \frac{2}{d}\*C_F\*C_A\*\frac{c_F^{(1)}}{m_1}\*\frac{c_F^{(2)}}{m_2}\*\frac{1}{\lvert \mathbf{k} \rvert} \*(d-1)
\*i\*I_d\right\}
\,,
\label{V1loopdEa} 
\end{align}
and
\begin{align}
 \langle  \delta \tilde V^{(2,b)}_{S^2,E}  \rangle_\text{HF}
=
\pi
\*\left(\frac{g_B^2}{4\pi}\right)^3(4\pi)^{2-d} {|\bf k|}^{2(d-3)}
\*C^2_F\*C_A\*\frac{c_F^{(1)}}{m_1}\*\frac{c_F^{(2)}}{m_2}
\frac{d-1}{d}
\frac{\Gamma(3-d)\Gamma^3(d/2-1)}{\Gamma(3d/2-3)}
\,.
\label{V1loopdEb}   
\end{align}

The first term generates an ${\cal O}(\als^2 /m^3)$ velocity-dependent potential, which we will discuss in conjunction with the genuine one-loop $1/m^3$ computation in Sec.~\ref{Sec:m3}. The second term generates an ${\cal O}(\als^3/m^2)$ potential, which we will discuss in conjunction with the genuine two-loop $1/m^2$ computation in the next Sec.~\ref{Sec:WL}.


\subsection{${\cal O}(\als^3/m^2)$ spin-dependent and velocity-independent potential with Wilson loops}
\label{Sec:WL}

The off-shell and Wilson-loop matching schemes are organized as expansions in powers of $1/m$.  Once the energy-dependent terms that may appear in the off-shell matching scheme have been eliminated, both matching schemes can be related through field redefinitions, which can themselves be organized in an expansion in powers of $1/m$ (see e.g. Refs.~\cite{Brambilla:2000gk,Peset:2015vvi}). Field redefinitions that depend only on $r$ commute with both the static potential and $V_{S^2}^{(1,1)}$. Therefore, they are identical in both matching schemes. This implies that $V_{S^2}^{(1,1)}$ can be computed in the Wilson-loop matching scheme and still be consistently combined with other potentials obtained using the off-shell matching scheme. In this paper, we have carried out this computation at two loops.

We emphasize that this discussion holds only after explicitly eliminating the energy-dependent terms in the pNRQCD Lagrangian. Otherwise, even leading-order potentials (the static potential for spin-independent observables and $V^{(1,1)}_{S^2}$ for the hyperfine splitting) may depend on the matching scheme. We remark that this applies even to the static potential, which would differ between the  off-shell and Wilson-loop matching schemes if energy-dependent terms are kept explicitly in the potential. 

 The determination of the $1/m^2$ spin-dependent and velocity-independent potential in the Wilson-loop matching scheme can be efficiently carried out using the same techniques employed for the determination of the static potential, as it exhibits a similar behavior (there are no velocity-dependent contributions). Therefore, an analogous exponentiation to that of the static potential holds, leading to the cancellation of Abelian-like terms.\footnote{An explicit realization of this idea for spin-independent and velocity-dependent potentials can be found in Ref.~\cite{Peset:2017wef}.} This links with the representation of $V_{S^2}^{(1,1)}$ in terms of Wilson loops  \cite{Eichten:1980mw,Pineda:2000sz}:
\begin{align}
     V^{(2)}_{S^2} \equiv \frac{V_{S^2}^{(1,1)}}{m_1m_2}{\bf S}_1\cdot {\bf S}_2
=
\frac{ c_F^{(1)} c_F^{(2)} }{ 6m_1m_2}i \lim_{T\rightarrow \infty}\int_0^{T} dt \,  
\lla g{\bf B}_1(t) \cdot g{\bf B}_2 (0) \rra
{\bfsigma}_1\cdot {\bfsigma}_2
+V_{\rm hard}
\,.
\end{align}
 $V_{\rm hard}$ is proportional to the Dirac delta and, therefore, does not contribute to P-wave states. Consequently, we will not consider it in this paper. We instead compute the soft contribution at two loops in momentum space and present the projected potential relevant for the hyperfine splitting. We obtain the following expression for the Fourier transform of $V^{(2)}_{S^2} $ (expressed in terms of the bare coupling $g_B$, the bare Wilson coefficients $c_F^{(i)}$, and $D=4-2\epsilon$)
\begin{align}
\label{VS2twoloops}
    &
\langle  \tilde V^{(2)}_{S^2} \rangle
_\text{HF}=
-\frac{8C_F\pi}{3m_1m_2}
\frac{g_B^2}{4\pi}
\frac{  c_F^{(1)}c_F^{(2)} (1-\epsilon ) }{1-\frac{2 \epsilon }{3}}
\\
\nn
&
\times
\left(
1+c_1\left( \frac{\mu^2}{{\bf k}^2} \right)^{\epsilon}
\frac{g_B^2\mu^{-2\epsilon}}{4\pi}
\frac{e^{-\gamma \epsilon}}{(4\pi)^{-\epsilon}}
 +c_2 \left( \frac{\mu^2}{{\bf k}^2} \right)^{2\epsilon}\left(\frac{g_B^2\mu^{-2\epsilon}}{4\pi}\right)^2 \frac{e^{-2\gamma \epsilon}}{(4\pi)^{-2\epsilon}}+\cdots
\right),
\end{align}
where
\begin{align}
c_1&=\frac{1}{4\pi}
\left[
\left( 
\frac{-\epsilon(\epsilon - 1) \Gamma(-\epsilon)^2 \Gamma(\epsilon)}{(8\epsilon^2 - 16\epsilon + 6)\Gamma(-2\epsilon)}
\cdot e^{\gamma \epsilon}
\cdot \left( C_A(4\epsilon - 5) + 4T_Fn_l \right)
\right)
\right.
\\
&
\nn
\left.
- \left(1 - \frac{2}{3} \epsilon\right)  3   \frac{\epsilon \Gamma(-\epsilon)^2 \Gamma(\epsilon)}{(8\epsilon - 4)\Gamma(-2\epsilon)} e^{\gamma \epsilon} C_A
\right].    
\end{align}

This full $D$ dependence has been taken from \cite{Beneke:2013jia}. The finite expression was computed in Ref.\ \cite{Gupta:1981pd}. 

The two-loop expression reads
\begin{align}
c_2&=\left( \frac{1 - \frac{2}{3} \epsilon}{1 - \epsilon} \right)
  \frac{\pi^{3}}{\left(3-2\,\epsilon\right)^{2}}\,\biggl(
 \nn \\
  &\frac{C_{A}^{2}}{\epsilon^2\,(1-2\,\epsilon)\,(3-2\,\epsilon)}\,\biggl[
\begin{aligned}[t]
  &+12\,(1-4\,\epsilon^2)\,(3-2\,\epsilon)^{2}\,\epsilon^{2}\,\centeredgraphics{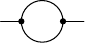}^2\biggr\rvert_{D \to d}\\
  &-96\,(1-2\,\epsilon)\,\epsilon\,\left(-3+39\,\epsilon-26\,\epsilon^{2}-10\,\epsilon^{3}+6\,\epsilon^{4}\right)\,\centeredgraphics{P1LML}^2\\
  &+32\,(1-2\,\epsilon)\,(3-2\,\epsilon)^{2}\,\epsilon^{2}\,\left(1+\epsilon\right)\,\centeredgraphics{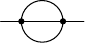}\biggr\rvert_{D \to d}\\
 &+384\,\left(-6+69\,\epsilon-280\,\epsilon^{2}+360\,\epsilon^{3}+176\,\epsilon^{4}
  \right.
  \\
  &
  \left.
  -853\,\epsilon^{5}+862\,\epsilon^{6}-396\,\epsilon^{7}+72\,\epsilon^{8}\right)\,\centeredgraphics{P2L3lML}\\
  &-96\,(3-2\,\epsilon)\,\epsilon\,\left(1-5\,\epsilon+38\,\epsilon^{2}-36\,\epsilon^{3}+8\,\epsilon^{4}\right)\,\centeredgraphics{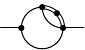}\biggr]\\
\end{aligned}\nn\\
&+\frac{C_{A}\,n_{l}\,T_{f}}{\epsilon\,(3-2\,\epsilon)}\,\biggl[
  \begin{aligned}[t]
    &-384\,\left(-6+19\,\epsilon-28\,\epsilon^{2}+25\,\epsilon^{3}-16\,\epsilon^{4}+4\,\epsilon^{5}\right)\,\centeredgraphics{P1LML}^{2}\\
   &-768\,\left(-34+203\,\epsilon-452\,\epsilon^{2}+481\,\epsilon^{3}-252\,\epsilon^{4}+52\,\epsilon^{5}\right)\,\centeredgraphics{P2L3lML}\\
    &-384\,(1-2\,\epsilon)\,(3-2\,\epsilon)\,\left(-2+\epsilon\right)\,\centeredgraphics{P2L4lML}\biggr]\\
  \end{aligned}\nn\\
 &+\left(\frac{1-\epsilon}{\epsilon}\right)^2\,C_{F}\,n_{l}\,T_{f}\,\biggl[
    \begin{aligned}[t]
     &-768\,\epsilon\,\left(2-\epsilon+2\,\epsilon^{2}\right)\,\centeredgraphics{P1LML}^{2}\\
      &+3072\,(1-2\,\epsilon)\,\left(2-2\,\epsilon+\epsilon^{2}\right)\,\centeredgraphics{P2L3lML}\biggr]\\
   \end{aligned}\nn\\
  &-\frac{3072\,(1-\epsilon)^{3}\,n_{l}^{2}\,T_{f}^{2}}{3-2\,\epsilon}\,\centeredgraphics{P1LML}^{2}\biggr),
\end{align}
where
\begin{align}
  \centeredgraphics{P1LML} ={} \frac{i}{(4\*\pi)^{D/2}}\*\frac{\Gamma\left(\frac{D}{2}-1\right)^2\*\Gamma\left(2-\frac{D}{2}\right)}{\Gamma(D-2)} \xrightarrow{D \to d} I_d,\\
  \centeredgraphics{P2L3lML} ={} \frac{1}{(4\*\pi)^{D}}\*\frac{\Gamma\left(\frac{D}{2}-1\right)^3\*\Gamma(3-D)}{\Gamma\left(\frac{3}{2}D-3\right)},\\
  \centeredgraphics{P2L4lML} ={} \frac{4^{4-D}}{(4\*\pi)^{D}}\*\Gamma\left(\frac{1}{2}\right)\*\frac{\Gamma\left(1+\frac{D}{2}\right)\*\Gamma(-D)\*\Gamma(-\frac{D}{2})\*\Gamma(1+D)}{\Gamma\left(D-\frac{5}{2}\right)}.
\end{align}
The single lines are standard scalar propagators, the double
line is a static propagator $1/(p_0 + i \eta)$, 
and the dot on the double
line indicates the square of the propagator. 

When expanded in $\epsilon$, we find 
\begin{align}
    &
\left( \frac{1 - \epsilon}{1 - \frac{2}{3} \epsilon} \right)
c_2
=
\frac{2C_A^2 - 17C_A\,n_l\,T_F + 8n_l^2\,T_F^2}{18\epsilon^2\pi}
\\
&
\nn
+ \frac{-7C_A^2 + 3C_A^2\pi^2 - 31C_A\,n_l\,T_F - 9C_F\,n_l\,T_F + 24n_l^2\,T_F^2}{18\epsilon\pi}
\\
&
\nn
+ \frac{1}{432\pi}
\Bigl(
-1150C_A^2 + 544C_A^2\pi^2 - 27C_A^2\pi^4 - 128C_A\,n_l\,T_F - 1908C_F\,n_l\,T_F
\\
&
\nn
+ 68C_A\,n_l\,\pi^2\,T_F + 1472n_l^2\,T_F^2 - 32n_l^2\,\pi^2\,T_F^2
+ 36C_A^2\zeta(3) - 2016C_A\,n_l\,T_F\,\zeta(3) 
\\
&
\nn
+ 1728C_F\,n_l\,T_F\,\zeta(3)
\Bigr)+
{\cal O}(\epsilon)
\,.
\end{align}

We can express the above result in terms of the renormalized coupling using the relation between the bare and the $\MS$-renormalized coupling:
 \begin{align}
 \frac{g_B^2\mu^{-2\epsilon}}{4\pi} \left(\frac{e^{\gamma }}{(4\pi)}\right)^{-\epsilon}
=
\als
 \left(
 1-
   \frac{\als 
   \text{$\beta_0$}}{4 \pi  \epsilon }+
 \frac{\als ^2 \text{$\beta_0$}^2}{16 \pi
   ^2 \epsilon ^2}-\frac{\als ^2 \text{$\beta_1 $}}{32 \pi ^2 \epsilon }\right)   
   \,,
 \end{align}
where $\beta_0=11/3C_A-4/3T_Fn_l$ and so on.

The hyperfine projection of the potential used in Eq.~\eqref{VS2twoloops} is sufficient if a single insertion of the potential appears in perturbation theory. This is not the case for the S-wave hyperfine splitting at N$^4$LO, where two insertions of the potential may appear in perturbation theory. In that case one has to keep track of the original Pauli-matrix dependence of the potential, but only for its leading order expression.

We have provided the $D$-dimensional expression for future reference. For the purposes of this paper, the four-dimensional expressions are sufficient, since the resulting potential is finite (up to terms proportional to the Dirac delta). It can be written as follows in terms of the renormalized $c_F$ in the $\MS$ scheme:
\begin{align}
\tilde V^{(2),\text{1-loop}}_{S^2,\text{P}} \equiv \frac{ \tilde V^{(1,1)}_{S^2}}{m_1m_2}{\bf S}_1\cdot {\bf S}_2\bigg |_\text{P-wave}
&\dot=
\frac{2\pi C_F}{3} \ln k \frac{c_F^\one c_F^\two}{m_1m_2} \frac{\als^2}{\pi}
\left(-\frac{\beta_0}{2}+\frac{7}{4}C_A\right)\bfsigma_1\cdot \bfsigma_2,
\label{V3HF}
\end{align}
at one loop. At two loops we have
\begin{align}&
\nn
\tilde V^{(2),\text{2-loop}}_{S^2,\text{P}} \equiv 
\frac{ \tilde V^{(1,1)}_{S^2}}{m_1m_2}{\bf S}_1\cdot {\bf S}_2\bigg |_\text{P-wave}
\\
\nn
&=\frac{ \als ^3 C_F c_F^{(1)} c_F^{(2)} }{27 \pi  m_1m_2}
{\bfsigma}_1\cdot {\bfsigma}_2
\Bigl[\left(2 C_A^2-17 C_A T_Fn_l+8 (T_Fn_l)^2\right) \left(\ln ^2k-2 \ln k \ln \mu \right)\nn\\
&-\frac{1}{3} \ln k \left(\left(1+9 \pi ^2\right) C_A^2-40 C_A T_Fn_l-27 C_F T_Fn_l+40 (T_Fn_l)^2\right)\Bigr]\label{VS22Finite}.
\end{align} 
For \cref{V3HF}, we need $c_F^{(i)}$ at NLL, which reads \cite{Amoros:1997rx,Czarnecki:1997dz}
\begin{align}\label{defcF}
c_F^{(i),\text{NLL}}&= z^{-\frac{\gamma_0}{ 2}}
    \left[ 1 + \frac{\alpha_s(\mu_h)}{4\pi}
      \left(c_1+\frac{\gamma_0}{2}\ln\frac{\mu_h^2}{m_i^2}\right) 
      + \frac{\alpha_s(\mu_h) - \alpha_s(\mu_s)}{4\pi}\left(
        \frac{\gamma_1}{2\beta_0} - \frac{\gamma_0\beta_1}{2\beta_0^2}
      \right) \right] ,
\end{align} 
where $c_1 = 2(C_A+C_F)$,
\begin{align}
\gamma_0 = 2 C_A \,, \qquad
   \gamma_1 = \frac{68}{9}\,C_A^2 - \frac{52}{9}\,C_A T_F\,n_l
\end{align}
and 
\begin{align}
    z=(\als(\mu)/\als(\mu_h))^{1/\beta_0} \label{defz}
    \,,
\end{align}
where $\mu$ and $\mu_h$ refer to the soft and hard scales respectively.

The $1/m^2$ spin-dependent and velocity-independent potential was also computed to two loops using the Wilson-loop matching scheme in Refs.~\cite{Kniehl:2003ap,Penin:2004ay,Penin:2004xi}. In those works, only the contributions proportional to $1/\epsilon^2$ and $1/\epsilon$ were considered as the aim was to compute the soft anomalous dimension. However, the ${\cal O}(\epsilon^0)$ term was also available. We have verified that our results agree with that computation to this order. Nevertheless, in those references, this potential was used together with the energy-dependent potential (\ref{VsoftE}), whereas the equivalence between off-shell and Wilson-loop scheme computations of the $1/m^2$ spin-dependent and velocity-independent potentials holds only after explicit energy-dependent terms have been eliminated via field redefinitions. This introduces additional corrections to the analyses in Refs.~\cite{Kniehl:2003ap,Penin:2004ay,Penin:2004xi}, which will be addressed in forthcoming papers.

We have also compared our result with the recent computation of Ref.~\cite{Mishima:2024afk}. This computation was performed in the on-shell matching scheme and the outcome was presented in an explicitly Hermitian form. We observe that, upon setting the hard contributions in that reference to zero and the $c_F$ Wilson coefficients to 1, their result coincides with ours.


\subsection{ One-loop potentials of order  $\als^2/m^3$}
\label{Sec:m3}

In this section, we compute the $1/m^3$ potential that contributes to the hyperfine splitting in general $D$ dimensions using the off-shell matching scheme. We only consider its projection to the hyperfine splitting.
We obtain
\begin{align}
    \langle  \delta \tilde V^{(3)}_{S^2}  \rangle_\text{HF}
=
    &
-
\frac{g_B^4}{16}\*\lvert{\bf k}\rvert^{d-3}
\*i\*I_d
\nn
\left\{
\frac{2}{d}\*C_F\*C_A\*\frac{c_F^{(1)}}{m_1}\*\frac{c_F^{(2)}}{m_2}\*\frac{1}{\lvert \mathbf{k} \rvert} \left((d-1)\*\frac{\mathbf{p}^2+\mathbf{p'}^2}{2\*m_r}-\frac{\mathbf{k}^2}{m_r}\right)
\right.
\\
\nn
    &
- \frac{1}{d}\*C_F\*[8\*(d-1)\*C_F - (2\*d - 3)\*C_A]\*\left(\frac{c_F^{(1)}}{m_1}\frac{c_S^{(2)}}{m_2^2} + \frac{c_S^{(1)}}{m_1^2}\frac{c_F^{(2)}}{m_2}\right)\* \lvert \mathbf{k} \rvert
    \\
&
\left.
+ \frac{(d-2)\*(d-4)}{d}\*C_F\*C_A\*\left(\frac{c_F^{(1)}}{m_1}\*\frac{c_F^{(2)2}}{m_2^2}+\frac{c_F^{(1)2}}{m_1^2}\*\frac{c_F^{(2)}}{m_2}\right) \*\lvert \mathbf{k} \rvert 
\right\},
\label{V1loopd}
\end{align}
where $I_d$ was defined in eq.~\eqref{Id}.

This quantity is relevant for N$^4$LO computations, as well as serving as a building block for N$^4$LL computations. If one is only interested in getting N$^3$LL accuracy for the S-wave hyperfine (for P-wave it does not contribute at N$^3$LL), the above result can be approximated to its four-dimensional expression:  
\begin{eqnarray}
\delta \tilde V^{(3)}_{S^2}&=&
-\frac{1}{ 24}(\pi\alpha_s)^2 C_F\left(4 C_F-\frac{3}{4}C_A\right)\left(\frac{c_F^{(1)}}{m_1}
\frac{c_S^{(2)}}{ m_2^2}+\frac{c_S^{(1)}}{ m_1^2}\frac{c_F^{(2)}}{m_2}
\right) |{\bfm k}|\bfm{\sigma}_1\cdot \bfm{\sigma}_2
\nn\\
&&
+\frac{1}{24}(\pi\alpha_s)^2 C_F C_A
\frac{c_F^{(1)}}{m_1}\frac{c_F^{(2)}}{m_2} \frac{1 }{ |{\bfm k}|}
\left(\frac{{\bfm p}^2+{\bfm p'}^2 }{ 2m_r}-\frac{{\bfm k}^2 }{ 2m_r}
\right)\bfm{\sigma}_1\cdot \bfm{\sigma}_2
\nn\\
&&
-\frac{1 }{ 96}(\pi\alpha_s)^2 C_FC_A\left(
\frac{c_F^{(1)}}{ m_1}\frac{c_F^{(2)2}}{ m_2^2}
+\frac{c_F^{(1)2}}{ m_1^2}\frac{c_F^{(2)}}{m_2}
\right) |{\bfm k}|\bfm{\sigma}_1\cdot \bfm{\sigma}_2\,.
\label{V1l}
\end{eqnarray}
This computation was done in Ref.~\cite{Penin:2004ay} (the Abelian piece also in Ref.~\cite{Czarnecki:1999mw}). Our expression confirms all the terms in eq.~(30) of \cite{Penin:2004ay}, except for the non-Abelian term proportional to $c_Fc_S$. This term has been corrected due to an extra diagram that was missing in the original computation. The consequences of this correction for the S-wave hyperfine splitting will be carried out elsewhere. 

To the above result one has to add  eq.~\eqref{V1loopdEa} to obtain the complete $\tilde V^{(3)}_{S^2}$, which reads 
\begin{align}
    \langle \tilde V^{(3)}_{S^2}  \rangle_\text{HF}
=
    &
-
\*\frac{g_B^4}{16} \lvert{\bf k}\rvert^{d-3}
\*i\*I_d
\nn
\left\{- \frac{2}{d}\*C_F\*C_A\*\frac{c_F^{(1)}}{m_1}\*\frac{c_F^{(2)}}{m_2}
     \frac{\lvert \mathbf{k} \rvert}{m_r}
\right.
\\
\nn
    &  
- \frac{1}{d}\*C_F\*[8\*(d-1)\*C_F - (2\*d - 3)\*C_A]\*\left(\frac{c_F^{(1)}}{m_1}\frac{c_S^{(2)}}{m_2^2} + \frac{c_S^{(1)}}{m_1^2}\frac{c_F^{(2)}}{m_2}\right)\* \lvert \mathbf{k} \rvert
    \\
&
\left.
+ \frac{(d-2)\*(d-4)}{d}\*C_F\*C_A\*\left(\frac{c_F^{(1)}}{m_1}\*\frac{c_F^{(2)2}}{m_2^2}+\frac{c_F^{(1)2}}{m_1^2}\*\frac{c_F^{(2)}}{m_2}\right) \*\lvert \mathbf{k} \rvert 
\right\}.
\label{V1loopdFinal}
\end{align}
We observe that the velocity-dependent term exactly cancels (for general $D$ dimensions). 

For the energy shift considered in this work, we will only need the three-dimensional expression of the potential above, which reads
\begin{align}
\tilde V_{S^2}^{(3)}&=-\frac{1}{24}(\pi\als)^2C_FC_A\frac{c_F^{(1)}c_F^{(2)}}{2m_r m_1m_2}k {\bfsigma}_1\cdot{\bfsigma}_2\nn\\
&-\frac{1}{24}(\pi\als)^2C_F\left(4C_F-\frac{3}{4}C_A\right)\left(\frac{c_F^{(1)}c_S^{(2)}}{m_1m_2^2}+\frac{c_F^{(2)}c_S^{(1)}}{m_2m_1^2}\right)k {\bfsigma}_1\cdot{\bfsigma}_2
\nn\\
&-\frac{1}{96}(\pi\als)^2C_FC_A\left(\frac{c_F^{(1)}c_F^{(2)2}}{m_1m_2^2}+\frac{c_F^{(2)}c_F^{(1)2}}{m_2m_1^2}\right)k {\bfsigma}_1\cdot{\bfsigma}_2.\label{VS23}
\end{align}
The computation of this one-loop $1/m^3$ potential has been carried out in the off-shell matching scheme, whereas the spin-dependent velocity-independent $1/m^2$ potential computed in Sec. \ref{Sec:WL} was obtained in the Wilson-loop matching scheme. Nevertheless, as discussed above, the latter is equivalent to the off-shell matching scheme, and both results can be safely combined. 

If we set the Wilson coefficients to one and $m_1=m_2$, the non-Abelian term proportional to $|{\bf k}|$ exactly agrees with the result in Ref.~\cite{Mishima:2024afk}, which was computed in the on-shell scheme. This term vanishes in four dimensions but not beyond (and one needs it beyond $D=4$ to reach N$^4$LO precision for S-wave observables). On the other hand, the abelian term is exactly 1/2 the corresponding term obtained in the on-shell matching scheme in Ref.~\cite{Mishima:2024afk}. This discrepancy is not problematic, as other potentials differ as well: i) the $1/m$ potential that enters at second-order perturbation theory is different in the on-shell and off-shell matching schemes and ii) there are additional potentials from the tree level computation at order $1/m^4$ that vanish on-shell. For the specific observable considered in this paper, we explicitly check in Sec. \ref{Sec:offon} that the N$^4$LO result (for equal masses) is identical, regardless of the matching scheme used.


\subsection{ Tree-level potentials of order $\als/m^4$}

We now consider the ${\cal O}(\als/m^4)$ tree-level potentials relevant for spin-dependent observables. For simplicity, we only display them in four dimensions.
\begin{align}
    \tilde V_{S^2,1}^{(4)}&=\pi  C_F\alpha_s {c_S^{(1)}c_S^{(2)}\over 4m_1^2m_2^2}
{1 \over{\bfm k}^2}\bfm{\sigma}_1\cdot({\bf k}\times{\bfm p})\bfm{\sigma}_2
\cdot({\bfm k}\times{\bfm p})\,,
\label{Vt1}\\
\tilde V_{S^2,2}^{(4)}&=-\pi C_F\alpha_s{c_F^{(1)}c_F^{(2)}\over
4m_1^2m_2^2}{({\bfm p}^2-{\bfm p'}^2)^2 \over {\bfm k}^2}
\left(\bfm{\sigma}_1\cdot \bfm{\sigma}_2
-{\bfm{\sigma}_1\cdot{\bfm k}\bfm{\sigma}_2\cdot{\bfm k}\over{\bfm k}^2}\right)\,,
\label{Vt2}\\
\tilde V_{S^2,3}^{(4)}&=-\pi  C_F\alpha_s {{\bfm p}^2-{\bfm p'}^2 \over 2{\bfm k}^2}
\left[{c_S^{(1)}c_F^{(2)} \over 4m_1^3m_2}
(\bfm{\sigma}_1\times ({\bfm p}+{\bfm p}'))\cdot(\bfm{\sigma}_2\times
{\bfm k})+(1\leftrightarrow 2)\right]\,,
\label{Vt3}\\
\tilde V_{S^2,4}^{(4)}&=-{\pi  C_F\alpha_s  \over 8{\bfm k}^2}
\left[{c_{pp'}^{(1)}c_F^{(2)} \over m_1^3m_2}\bfm{\sigma}_1\cdot ({\bfm p}+{\bfm p}')
\left(\bfm{\sigma}_2\cdot({\bfm p}+{\bfm p}') {\bfm k}^2-({\bfm p}^2-{\bfm p'}^2)
\bfm{\sigma}_2\cdot {\bfm k}\right)\right.
\nn\\
&
+(1\leftrightarrow 2)\Bigg]\,,
\label{Vt4}
\\
\nn
\tilde V^{(4)}_{S^2,5}&=
-\frac{C_F\pi\als}{4}\frac{{\bf p}^2+{\bf p}'^2}{{\bf k}^2}
\left(
\frac{c^{(1)}_{W_1}c_F^{(2)}}{m_1^3m_2}
+
\frac{c^{(2)}_{W_1}c_F^{(1)}}{m_2^3m_1}
\right)
(\bfm{\sigma}_1\times {\bf k})\cdot (\bfm{\sigma}_2 \times {\bf k})
\\
&=
\frac{C_F\pi\als}{16}\frac{{\bf p}^2+{\bf p}'^2}{{\bf k}^2}
\left(
\frac{c^{(1)}_{W_1}c_F^{(2)}}{m_1^3m_2}
+
\frac{c^{(2)}_{W_1}c_F^{(1)}}{m_2^3m_1}
\right)
[\bfm{\sigma}_1\cdot {\bf k},\bfm{\sigma}^n_1]
[\bfm{\sigma}_2\cdot {\bf k},\bfm{\sigma}^n_2]
\,,
\label{Vt5}\\
\tilde V^{(4)}_{S^2,6}&=\frac{C_F\pi\als}{2}\frac{{\bf p}\cdot{\bf p}'}{{\bf k}^2}
\left(
\frac{c^{(1)}_{W_2}c_F^{(2)}}{m_1^3m_2}
+
\frac{c^{(2)}_{W_2}c_F^{(1)}}{m_2^3m_1}
\right)
(\bfm{\sigma}_1\times {\bf k})\cdot (\bfm{\sigma}_2 \times {\bf k})
\nn\\
&=
-\frac{C_F\pi\als}{8}\frac{{\bf p}\cdot{\bf p}'}{{\bf k}^2}
\left(
\frac{c^{(1)}_{W_2}c_F^{(2)}}{m_1^3m_2}
+
\frac{c^{(2)}_{W_2}c_F^{(1)}}{m_2^3m_1}
\right)
[\bfm{\sigma}_1\cdot {\bf k},\bfm{\sigma}^n_1]
[\bfm{\sigma}_2\cdot {\bf k},\bfm{\sigma}^n_2]
\,,
\label{Vt6}
\\
\tilde V^{(4)}_{S^2,7}&=\frac{iC_Fg^2}{16m_1^2 m_2^2}
[c_D^{(1)}c_S^{(2)}+c_S^{(1)}c_D^{(2)}]
{\bf S} \cdot ({\bf p} \times {\bf k})
\,.
\label{Vt7}
\end{align}

Here $\bfm p$ and $\bfm{p}'$ are the momentum of
incoming and outgoing quark, whereas ${\bf k}={\bf p}-{\bf p}'$.  

The first four potentials were already computed in Ref.~\cite{Penin:2004ay} in the off-shell matching scheme (after eliminating any energy-dependent terms). Here we only add those that are relevant for the computation of this paper or for future computations of spin-dependent observables.  

There are  ${\cal O}(v^4)$ potentials involving the product of
$c_{W_2}c_F$ or $c_{W_1}c_F$, which are proportional to $({\bfm
  p}\cdot {\bfm p'})(\bfm{\sigma}_1 \cdot \bfm{\sigma}_2)$ and $({\bfm
  p}^2+{\bfm p'}^2)(\bfm{\sigma}_1 \cdot \bfm{\sigma}_2)$, respectively. Consequently, these potentials do not generate divergences at N$^3$LL, but they may contribute at N$^4$LO and N$^4$LL. Since $c_{W_2}$ is ${\cal O}(\als)$, it does not contribute to N$^4$LO. Nevertheless, it can contribute at N$^4$LL. Therefore, we compute the potentials proportional to these Wilson coefficients, which yield the potentials $\tilde V^{(4)}_{S^2,5/6}$ in \cref{Vt5,Vt6}. Finally, another potential must be considered, at least for P-wave states. This potential is proportional to $c_D\times c_S$, which we provide in $\tilde V^{(4)}_{S^2,7}$ in \cref{Vt7}. 

As mentioned above, by using the commutation relations of the Pauli matrices, we have written the above potentials in four dimensions. We could as well have left them unevaluated, which is convenient when computing traces in $D$ dimensions. This is necessary only for divergent loops, which is not the case of this paper. If needed, it is straightforward to keep track of the full Pauli-matrix dependence. We have illustrated this explicitly for the potentials $\tilde V^{(4)}_{S^2,5}$ and $\tilde V^{(4)}_{S^2,6}$ in \cref{Vt5,Vt6}. 

 Another point worth emphasizing is that some of these potentials generate quasi-local terms, proportional to the Dirac delta or derivatives of it. This becomes apparent when the potentials are written in position space. For instance, for $V^{(4)}_{S^2,1}$ and $V^{(4)}_{S^2,7}$, we have (these potentials agree with the analogous QED expressions once setting the Wilson coefficients to one \cite{Zatorski})
\begin{align}
    V^{(4)}_{S^2,1}
=
\frac{c_S^{(1)}c_S^{(2)}C_Fg^2}{4m_1^2m_2^2}
({\bf S}_1 \times {\bf p})^i 
\left\{
\frac{1}{r^3}\left(\delta^{ij}-3\frac{r^ir^j}{r^2}\right)
+\frac{\delta^{ij}}{3}(4\pi)\delta({\bf r})
\right\}
({\bf S}_2 \times {\bf p})^j
\,,
\end{align}
\begin{align}
  V^{(4)}_{S^2,7}
=
\frac{iC_Fg^2}{16m_1^2 m_2^2}
[c_D^{(1)}c_S^{(2)}+c_S^{(1)}c_D^{(2)}]S^i p^n\delta({\bf r}) p^m \epsilon^{inm}  
\,.
\end{align}

This fact indicates a possible mixing of these potentials with those generated by dimension-eight four-fermion operators in the NRQCD Lagrangian (for explicit expressions, see \cite{Bodwin:1994jh}). While this is not an issue for $V^{(4)}_{S^2,7}$, which we do not need in this paper, it is relevant for $V^{(4)}_{S^2,5}$ and $V^{(4)}_{S^2,6}$ as we now explain.

  For the $\als/m^4$ potentials, the NRQCD Wilson coefficients are required with LL accuracy to achieve N$^4$LL precision. Except for $c_{W_i}$, they can be rewritten in terms of $c_F$ using reparameterization invariance \cite{Manohar:1997qy}: 
  \begin{align}
  c_{S}=2c_{F}-1, \qquad c_{p'p}=c_F-1,
\end{align}
 whereas for $c_{W_i}$ the following relation holds
\begin{align}
    c_{W_1}=1+c_{W_2}.
\end{align}
In Ref.~\cite{Manohar:1997qy}, these coefficients were computed to one loop in Feynman gauge, and in Refs.\ \cite{Lobregat:2018tmn,Moreno:2018lbo} their leading-logarithmic renormalization-group running was obtained in the Coulomb gauge. 

The implementation of effects proportional to $c_{W_1}$ and $c_{W_2}$ is delicate because, as discussed in Ref.~\cite{Lobregat:2018tmn}, these Wilson coefficients are gauge dependent. Only the combination $c_{W_1}-c_{W_2}=1$ is gauge independent. This is the combination that appears in physical observables involving a single heavy quark. In the case of two heavy quarks, $c_{W_1}$ and $c_{W_2}$ may appear separately, but always in combination with the Wilson coefficients of the four-fermion operators. A similar situation occurs for the Wilson coefficients $c_D$ and $d_{vs}$, as discussed in Ref.~\cite{Pineda:2001ra}. In view of this, we rewrite the potentials $\tilde V^{(4)}_{S^2,5}$ and $\tilde V^{(4)}_{S^2,6}$ in the following way:
\begin{align}
    \tilde V^{(4)}_{S^2,5'}&=
-\frac{C_F\pi\als}{4}\frac{{\bf p}^2+{\bf p}'^2}{{\bf k}^2}
\left(
\frac{c_F^{(2)}}{m_1^3m_2}
+
\frac{c_F^{(1)}}{m_2^3m_1}
\right)
(\bfm{\sigma}_1\times {\bf k})\cdot (\bfm{\sigma}_2 \times {\bf k})
\,,
\label{Vt5prime}\\
\tilde V^{(4)}_{S^2,6'}&=\frac{C_F\pi\als}{4}
\left(
\frac{c^{(1)}_{W_2}c_F^{(2)}}{m_1^3m_2}
+
\frac{c^{(2)}_{W_2}c_F^{(1)}}{m_2^3m_1}
\right)
(\bfm{\sigma}_1\times {\bf k})\cdot (\bfm{\sigma}_2 \times {\bf k})
\,.
\label{Vt6prime}
\end{align}
This makes it explicit that the second potential is local.

Finally, we compare the set of potentials computed in this section with those obtained in the on-shell matching scheme with leading-order NRQCD Wilson coefficients (see \cite{Mishima:2024afk}). We observe that our potentials differ from the on-shell results. However, if we impose the on-shell condition $\vp'^2=\vp^2$ on the tree-level $1/m^4$ potential computed above, we recover the on-shell potential. 


\section{Quantum-mechanical perturbation theory}
\label{Sec:Delta}

With the results for the potential obtained above, we are now in a position to compute the hyperfine splitting of the P-wave heavy quarkonium spectrum with N$^4$LO precision, ${\cal O}(m\als^6)$, and (partially) with N$^4$LL precision, ${\cal O}(m\als^6)\times (\als \ln)^n$. The spectrum can be computed using standard quantum-mechanical perturbation theory. Note that there are no ultrasoft contributions to the hyperfine splitting up to ${\cal O}(m\als^6)$. 

Since we are computing hyperfine splittings, it is useful to employ the following relation (in four dimensions):
\begin{align}
    \langle {\bfsigma}_1^i{\bfsigma}_2^j\rangle_\text{HF}=-\frac{4}{3}\delta^{ij}\label{HFspinproj}
\,.
\end{align}

Here, we quote only the terms in the potential at each order that contribute to N$^4$LO.

From the static potential, we require only
\begin{align}\label{defV1}
V_1&=-\frac{C_F\alpha_s}{r}\frac{\als}{4\pi}\left(a_1+2\beta_0\ln(\mu r e^{\gamma_E})\right)
\end{align}
where $a_1=(31C_A- 20T_Fn_l)/36$.

From the tree-level $1/m^2$ potential, we require 
\begin{align}\label{defV2F}
 V_2^\text{F}&=\frac{\als  C_F  }{2 m_1 m_2 }\left(  c_F^{(1)}+ c_F^{(2)} +c_S^{(1)} \frac{m_2}{2m_1}+c_S^{(2)} \frac{m_1}{2m_2}\right)\frac{{\bf L\cdot S}}{r^3}+\frac{\als  C_F }{4 m_1 m_2 }c_F^{(1)} c_F^{(2)}\frac{{\bf S}_{12}}{r^3}\nn\\
&+\frac{\als  C_F }{2 m_1 m_2 }\left( c_F^{(1)}- c_F^{(2)} +c_S^{(1)} \frac{m_2}{2m_1}-c_S^{(2)} \frac{m_1}{2m_2}\right)\frac{{\bf L\cdot S^-}}{r^3}.
\end{align}
where $S^-=S_1-S_2$.  Note that the last operator does not contribute to $\Delta$.

From the $1/m^2$ spin-dependent and velocity-independent potential we require the one-loop, eq. (\ref{V3HF}), and two-loop, eq. (\ref{VS22Finite}), expressions. 

From the $1/m^3$ spin-dependent one-loop potential we only need its four-dimensional expression, \cref{VS23}.

To compute the energy, we then define
\begin{align}\label{deftV3HFP}
V_3^{S^2,\text{P}}&=V_{S^2,P}^{(2),\text{1-loop}},
\end{align}
where the right-hand side of the equation is given in \cref{V3HF}, and 
\begin{align}
\tilde{V}_4^{S^2,\text{P}}=\sum_{i=1}^6\tilde V_{S^2,i}^{(4)}+\tilde V_{S^2}^{(3)}+\tilde V_{S^2}^{(2),\text{2-loop}},
\label{deftV4HFP}
\end{align}
where each term is defined in \cref{Vt1,Vt2,Vt3,Vt4,Vt5,Vt6,VS23,VS22Finite}. 
In this notation $V_i$ stands for all the potentials that produce an $\mathcal{O}(m\als^{2+i})$ contribution to the energy levels with LO Wilson coefficients.

Note that operators such as $ -\frac{2 \pi  \als  C_F  }{3 m^4}\left(p^2+p'^2\right){\bf S}^2$, $\frac{3 \pi  \als  C_F }{m^4}\left(p^2+p'^2\right)\Lambda(\vk)$ and $\frac{ \pi  \als  C_F }{2m^4}\vk^2\Lambda(\vk)$ with $\Lambda(\vk)=i{\bf S}(\vp\times\vk)/\vk^2$, do not contribute to the P-wave hyperfine splitting.

We can now compute $\Delta$. We find at LO
\begin{align}
\label{EnHFPatN3LO}
    \Delta^{\text{LO}}&=\langle V_3^{S^2,\text{P}}\rangle_\text{UF}
\end{align}
and, at NLO,
\begin{align}\label{EnHFPatN4LO}
\Delta^{\text{NLO}}&=\langle V_4^{S^2,\text{P}}\rangle_\text{UF}-2\langle V_3^{S^2,\text{P}} \overline{G}_n V_1\rangle_\text{UF}- \langle  V_2^{\text{F}} \bar{G}_n V_2^{\text{F}}\rangle_\text{UF},
\end{align}  
where 
$$
\overline{G}_n=\sum_{m\not=n}\frac{|m\rangle\langle m|}{E_m-E_n}
$$ 
is the reduced Green's function. 

\subsection{First-order perturbation theory}
Using eq.~\eqref{HFspinproj} for an $n_rP$ state (recall that $n_r=n-1$) we find for the $\mathcal{O}(\als/m^4)$ potentials
\begin{align}
\langle V_{S^2,1}^{(4)}\rangle_\text{UF} &=\frac{\pi  \als  C_F c_S^{(1)}c_S^{(2)} }{12  m_1^2m_2^2}\langle\vk^2-2  \left(\vp^2+\vp'^2\right)+\frac{\left(\vp^2-\vp'^2\right)^2}{\vk^2}\rangle=-\frac{\als ^6 C_F^6 c_S^{(1)} c_S^{(2)} m_r^5 \left(2 n^2-3\right)}{90 m_1^2 m_2^2 n^5},\\
\langle V_{S^2,2}^{(4)}\rangle_\text{UF} &=\frac{2 \pi  \als   C_F c_F^{(1)}c_F^{(2)}}{3  m_1^2m_2^2}\langle\frac{\left(\vp^2-\vp'^2\right)^2}{\vk^2}\rangle=\frac{4 \als ^6 C_F^6 c_F^{(1)} c_F^{(2)} m_r^5 \left(3 n^2-2\right)}{45 m_1^2 m_2^2 n^5},\\
\langle V_{S^2,3}^{(4)}\rangle_\text{UF} &=\frac{ \pi  \als  C_F \left(c_F^{(1)}c_S^{(2)}\frac{m_1}{m_2}+c_F^{(2)}c_S^{(1)}\frac{m_2}{m_1}\right)}{3  m_1^2m_2^2}\langle\frac{\left(\vp^2-\vp'^2\right)^2}{\vk^2}\rangle\nn\\
&=\frac{2 \als ^6 C_F^6 m_r^5  \left(c_F^{(1)} c_S^{(2)} m_1^2+c_F^{(2)} c_S^{(1)} m_2^2\right)\left(3 n^2-2\right)}{45 m_1^3 m_2^3 n^5},
\end{align}
\begin{align}
\langle V_{S^2,4}^{(4)}\rangle_\text{UF} &=-\frac{\pi  \als  C_F \left(c_F^{(1)} c_{pp'}^{(2)}\frac{m_1}{m_2}+c_F^{(2)} c_{pp'}^{(1)}\frac{m_2}{m_1}\right) }{6  m_1^2m_2^2}\langle\vk^2-2  \left(\vp^2+\vp'^2\right)+\frac{\left(\vp^2-\vp'^2\right)^2}{\vk^2}\rangle\nn\\
&=\frac{\als ^6 C_F^6 m_r^5  \left(c_F^{(1)} c_{pp'}^{(2)} m_1^2+c_F^{(2)} c_{pp'}^{(1)} m_2^2\right)\left(2 n^2-3\right)}{45 m_1^3 m_2^3 n^5}\,\\
\langle V_{S^2,5'}^{(4)}\rangle_\text{UF} &=\frac{2 \pi  \als  C_F \left(c_F^{(1)} \frac{m_1}{m_2}+c_F^{(2)} \frac{m_2}{m_1}\right)}{3 m_1^2m_2^2}\langle\vp^2+\vp'^2\rangle=0,\\
\langle V_{S^2,6'}^{(4)}\rangle_\text{UF}&=-\frac{ 2\pi  \als  C_F\left(c_F^{(1)} c_{W_2}^{(2)}\frac{m_1}{m_2}+c_F^{(2)} c_{W_2}^{(1)}\frac{m_2}{m_1}\right) }{3 m_1^2m_2^2}\langle -\vk^2\rangle\nn\\
&=-\frac{4 \als ^6 C_F^6 m_r^5  \left(c_F^{(1)} c_{W_2}^{(2)} m_1^2+c_F^{(2)} c_{W_2}^{(1)} m_2^2\right)\left(n^2-1\right)}{9 m_1^3 m_2^3 n^5}.
\end{align}

One should also compute the expectation value of the ${\cal O}(\alpha_s^2/m^3)$ term. For this term, as well as the preceding one, one may set $c_F = 1$ if only the fixed-order computation is of interest:
\begin{align}
&
\langle V_{S^2}^{(3)}\rangle_\text{UF} =\frac{\pi ^2 \als ^2 C_F }{24 m_1^2 m_2^2 } \left[16 C_F  (c_F^{(1)} c_S^{(2)} m_1+c_F^{(2)} c_S^{(1)} m_2)\right.\nn\\
&\left.+C_A \left(c_F^{(1)2} c_F^{(2)} m_2 + c_F^{(2)2}c_F^{(1)} m_1 +2 c_F^{(2)} c_F^{(1)} (m_1+m_2)-3 c_S^{(2)}c_F^{(1)} m_1 -3 c_F^{(2)} c_S^{(1)} m_2 \right)\right]\langle k\rangle\nn\\
&=\frac{ m_r^4\als ^6 C_F^5(2-3n^2) }{180 m_1^2 m_2^2 n^5 } \left[16 C_F  (c_F^{(1)} c_S^{(2)} m_1+c_F^{(2)} c_S^{(1)} m_2)\right.\nn\\
&\left.+C_A \left(c_F^{(1)2} c_F^{(2)} m_2 + c_F^{(2)2}c_F^{(1)} m_1 +2 c_F^{(2)} c_F^{(1)} (m_1+m_2)-3 c_S^{(2)}c_F^{(1)} m_1 -3 c_F^{(2)} c_S^{(1)} m_2 \right)\right].
\end{align}
On top of that, we need to consider the logarithmic corrections to the $1/m^2$ potential. The expectation value of the ${\cal O}(\als^2/m^2)$ and ${\cal O}(\als^3/m^2)$ terms reads:
\begin{align}
&
\langle V_{S^2}^{(2)}\rangle_\text{UF}=\frac{8\pi C_F}{3} \frac{c_F^\one c_F^\two}{m_1m_2} \frac{\als^2}{\pi}
\left(-\frac{\beta_0}{2}+\frac{7}{4}C_A\right)\langle \ln k \rangle\nn\\
&-\frac{4 \als ^3 C_F c_F^{(1)} c_F^{(2)}  }{27 \pi  m_1m_2}\left[\left(2 C_A^2-17 C_A T_Fn_l+8 T_F^2n_l^2\right) \left(\langle\ln ^2k\rangle-2 \langle\ln k\rangle \ln \mu \right)\right.\nn\\
&
\left.-\frac{1}{3} \langle\ln k\rangle \left(\left(1+9 \pi ^2\right) C_A^2-40 C_A T_Fn_l-27 C_F T_Fn_l+40 T_F^2n_l^2\right)\right]\nn\\
&=-\frac{ m_r^3\alpha_s^5 C_F^4 c_F^{(1)} c_F^{(2)} (C_A-8 T_F n_l)}{54 \pi  m_1 m_2 n^3}\nn\\
&-\frac{m_r^3\als ^6 C_F^4 c_F^{(1)} c_F^{(2)}   }{486 \pi ^2 m_1 m_2 n^4}\left[-C_A (301 n+306) T_F n_l-54 C_F n T_F n_l+8 (23 n+18) T_F^2 n_l^2\right.\nn\\
&+\left.2 C_A^2 \left(\left(14+9 \pi ^2\right) n+18\right)+12 n (2 C_A-T_F n_l) (C_A-8 T_F n_l) \left(\ln \frac{a\mu  n}{2 }-H_{n+1}\right)\right],
\end{align}
where $a=1/(m_rC_F\als)$ is the Bohr radius and $H_n = \sum_{k=1}^n \frac{1}{n}$ the $n$th harmonic number. The ${\cal O}(\als^6)$ logarithmic term was already obtained in eq.~(4.9) of Ref.~\cite{Peset:2018jkf}.

\subsection{Second-order perturbation theory}

We compute the contributions from second-order perturbation theory using the formulas provided in Appendix~\ref{App:ME}. We first have
\begin{align}
-2&\langle V_3^{S^2,\text{P}}\overline{G}_nV_1\rangle_\text{UF} =\frac{\als ^6 C_F^4 m_r^3 (C_A-8 T_Fn_l)c_F^{(1)}c_F^{(2)} }{108 m_1m_2\pi ^2 n^4 }
\nn
\\
&
\times
\left[ \beta_0 n  \left(-3 \ln \left(\frac{a \mu  n}{2}\right)-H_{n-2}+2 n \psi ^{(1)}(n-1)-\frac{23}{6}-\frac{9 n^4-4 n^2+3}{n\left(n^2-1\right)^2}\right)+4 C_A n  \right],
\end{align}
where $a=1/(m_rC_F\als)$ is the Bohr radius, $H_n = \sum_{k=1}^n \frac{1}{n}$ the $n$th harmonic number, and $\psi^{(1)}$ the first-order polygamma function. 

We now consider the iteration of two relativistic ($1/m$) potentials. 
Note that the double insertion of the potential $ V_2^\text{F}$ acquires an off-diagonal contribution
\begin{align}
-\langle V_2^\text{F}\overline{G}_nV_2^\text{F}\rangle_\text{UF}&=-\left(\frac{3  \als  C_F }{2m^2}\langle {\bf L\cdot S}\rangle+\frac{  \als  C_F }{ 4m^2}\langle
{\bf S}_{12}\rangle\right)^2\langle \frac{1}{r^3}\overline{G}_n\frac{1}{r^3}\rangle+\delta E_{V_2,V_2}^\text{off},
\end{align}
where for an $nP$ state
\begin{align}
\delta E_{V_2,V_2}^\text{off}&=-\left(-\frac{3 \als  C_F c_F^{(1)}c_F^{(2)}}{2 m_1m_2}\right)^2|\langle n,l=1| (\frac{{\bf S\cdot r}}{r})^2|n,l=3|^2\nn\\
&\times\langle n,l=1| \frac{1}{r^3} \sum_{m\neq n}\frac{|m ,l=3\rangle\langle m,l= 3|}{E_m-E_n}\frac{1}{r^3}|n,l=1\rangle \nn\\
&=32m_r\left(-\frac{3 m_r^2\als^3  C_F^3 c_F^{(1)}c_F^{(2)}}{2 m_1m_2 }\right)^2|\xi_3(j,1,s)|^2G_{1,3}\nn\\
&=\frac{ m_r^5\als ^6 C_F^6}{1620 m_1^2 m_2^2 n^5}\frac{18}{5} c_F^{(1)2} c_F^{(2)2} \left(10 n^2-7\right)
\end{align}
where $G_{L,L+2}$ and $\xi_{L+2}(j,l,s)$ are defined in eqs.~(183) and~(197) of \cite{Zatorski} respectively. Note that for $G_{L,L+2}$ there is a typo regarding the sign of $c_1$ where it should be $c_1=\frac{1}{12 (l+1) (l+2) (2 l+3)}$.

For the P-wave hyperfine splitting we then find:
\begin{align}
&
-\langle V_2^\text{F}\overline{G}_nV_2^\text{F}\rangle_\text{UF}=\frac{ m_r^5\als ^6 C_F^6}{1620 m_1^2 m_2^2 n^5}\left[\frac{18}{5} c_F^{(1)2} c_F^{(2)2} \left(10 n^2-7\right)\right.\nn\\
&\left.+\left(227 n^2+90 n-108\right) \left(\left(c_F^{(1)}+c_F^{(2)}+c_S^{(1)} \frac{m_2}{2 m_1}+c_S^{(2)}\frac{ m_1}{2 m_2}\right)^2+\frac{3 }{5}c_F^{(1)2} c_F^{(2)2}\right)\right].
\end{align}

\section{Final result}
\label{Sec:FunalResult}

Altogether, for the $2P$ hyperfine splitting, we obtain 
\begin{align}
&\Delta=\frac{ m_r^3\alpha_s^5 C_F^4 c_F^{(1)} c_F^{(2)} (8 T_F n_l-C_A)}{54 \pi  m_1 m_2 n^3}\nn\\
&+\frac{m_r^3 \alpha_s^6 C_F^4 }{486 \pi ^2 m_1 m_2 n^3}\left[c_F^{(1)} c_F^{(2)} \left(54 C_F T_F n_l-372 C_A T_F n_l-\frac{3}{4}  \left(24 \pi ^2-71\right) C_A^2\right.\right.\nn\\
&\left.\left.-\frac{9}{2} (8 T_F n_l-C_A) \left[2 C_A \left(\ln \left(\frac{\mu  n}{2 \alpha_s  C_F m_r}\right)-H_{n+1}+\frac{3}{2 n}\right)\right.\right.\right.\nn\\
&\left.\left.\left.-\beta_0 \left(5\ln \left(\frac{\mu  n}{2 \alpha_s  C_F m_r}\right)-5H_{n-2}+4 H_{n+1}-2 n \psi ^{(1)}(n-1)-\frac{7}{n}+\frac{23}{3}+\frac{ n^4+7}{n \left(n^2-1\right)^2}\right)\right]\right)\right.\nn\\
&\left.-\frac{27 C_FC_A\pi ^2   \left(3 n^2-2\right)  }{10 n^2}\left(2 c_F^{(1)} c_F^{(2)}+\frac{c_F^{(1)2} c_F^{(2)}m_r}{m_1}+\frac{c_F^{(1)} c_F^{(2)2}m_r}{m_2}-\frac{3 c_F^{(1)} c_S^{(2)}m_r}{m_2}-\frac{3 c_F^{(2)} c_S^{(1)}m_r}{m_1}\right)\right.\nn\\
\nn
&+\frac{54  C_F^2 \pi ^2 m_r^2}{5 m_1 m_2 n^2} \left[\frac{1}{2}c_F^{(1)2} c_F^{(2)2}\left(-5+3n+\frac{287n^2}{30}\right)
\right.
\\
\nn
&
\qquad
+2 \left(2 c_F^{(1)} c_F^{(2)}- c_F^{(1)} c_S^{(2)} \left(\frac{m_1}{m_r}+1\right)- c_F^{(2)} c_S^{(1)} \left(\frac{m_2}{m_r}+1\right)\right)\left(-2+3 n^2\right)\nn\\
&
\qquad
\left.\left.+ \left(c_F^{(1)}+c_F^{(2)}+\frac{c_S^{(1)} m_2}{2 m_1}+\frac{c_S^{(2)} m_1}{2 m_2}\right)^2\left(-3+\frac{5n}{2}+\frac{227n^2}{36}\right)\right.\right.\nn\\
&
\qquad
\left.\left.+20 \left(\frac{c_{W_2}^{(2)} c_F^{(1)} m_1}{m_2}+\frac{c_{W_2}^{(1)} c_F^{(2)} m_2}{m_1}\right)\left(1-n^2\right) \right.\right.\nn\\
&
\qquad
\left.\left.+ \left(\frac{c_F^{(1)} c_{pp'}^{(2)} m_1}{m_2}+\frac{c_F^{(2)} c_{pp'}^{(1)} m_2}{m_1}-\frac{c_S^{(1)} c_S^{(2)}}{2}\right)\left(-3+2 n^2\right)\right]\right].
\end{align}
This expression is correct to N$^4$LO. It also incorporates the large hard logarithms arising from the Wilson coefficients of the NRQCD bilinear terms that appear at N$^4$LL. Note that to reach such precision, we need $c_F^{(i)}$ at NLL and all the other Wilson coefficients at LL. 

For ease of reference, we also provide the strict N$^4$LO expression, which reads
\begin{align}
&
\Delta^\text{N$^4$LO}=\frac{m_r^3\als ^5 C_F^4 (8 T_Fn_l-C_A) }{54 \pi  m_1 m_2 n^3}\nn\\
&+\frac{m_r^3\als^6 C_F^4  }{243 \pi ^2 m_1 m_2 n^3}\left[27 C_F T_Fn_l-150 C_A T_Fn_l-\frac{3}{8}  \left(24 \pi ^2-59\right) C_A^2\right.\nn\\
&\left.-\frac{9}{4} (8 T_Fn_l-C_A) \left(-2 C_F+2 C_A \left(\ln \left(\frac{n \sqrt{m_1 m_2}}{2 \alpha  C_F m_r}\right)-H_{n+1}+\frac{3}{2 n}\right)\right.\right.\nn\\
&\left.\left.-\beta_0 \left(5\ln \left(\frac{\mu  n}{2 \alpha  C_F m_r}\right)-5H_{n-2}+4 H_{n+1}-2 n \psi ^{(1)}(n-1)-\frac{7}{n}+\frac{23}{3}+\frac{ n^4+7}{n \left(n^2-1\right)^2}\right)\right)\right.\nn\\
&\left.+\frac{27 \pi ^2 C_F^2 }{5 n^2}\left(1+\frac{5 n}{2}+\frac{11 n^2}{36}-\frac{4m_r^2}{m_1m_2}\left(1-n-\frac{227 n^2}{90}\right)-\frac{m_1m_2}{4 m_r^2}\left(3-\frac{5 n}{2}-\frac{227 n^2}{36}\right) \right)\right].
\end{align}

The results for equal masses are obtained simply by setting $m_1=m_2=m$ and read
\begin{align}
    \Delta&=\frac{m \als ^5 C_F^4}{432 \pi  n^3} c_F^2 (8 T_Fn_l-C_A)\nn\\
    &+\frac{m\als ^6 C_F^4  }{3888 \pi ^2 n^3}\left[c_F^2 \left(54 C_F T_Fn_l-372 C_A T_Fn_l-\frac{3}{4}  \left(24 \pi ^2-71\right) C_A^2+\right.\right.\nn\\
    &\left.\left.-\frac{9}{2} (8 T_Fn_l-C_A) \left(2 C_A \left(\ln \frac{\mu  n}{\alpha  C_F m}-H_{n+1}+\frac{3}{2 n}\right)\right.\right.\right.\nn\\
    &\left.\left.\left.-\beta_0 \left(5\ln \frac{\mu  n}{\alpha  C_F m}-5H_{n-2}+4 H_{n+1}-2 n \psi ^{(1)}(n-1)-\frac{7}{n}+\frac{23}{3}+\frac{ n^4+7}{n \left(n^2-1\right)^2}\right)\right)\right)\right.\nn\\
    &\left.-\frac{ 27 \pi ^2 C_F C_A  }{10 n^2}\left(c_F^3+2 c_F^2-3 c_F c_S\right)\left(-2+3 n^2\right)\right.\nn\\
    &\left.+\frac{27 \pi ^2 C_F^2 }{5 n^2}\left[\frac{1}{4}c_F^4  \left(-5+3n+\frac{287n^2}{30}\right)+c_F^2  \left(-10+5n+\frac{335n^2}{18}\right)\right.\right.\nn\\
    &\left.\left.
    \qquad+c_Fc_S \left(6+5n-\frac{97n^2}{18}\right)+\frac{1}{4}c_S^2 \left(-3+5n+\frac{191n^2}{18}\right)\right.\right.\nn\\
    &\left.\left.
    \qquad+20c_Fc_{W_2}  \left(1-n^2\right)+c_F c_{pp'}  \left(-3+2 n^2\right)\right.\bigg]\right.\bigg]
    ,
\end{align}
and, at N$^4$LO,
\begin{align}
    \Delta^\text{N$^4$LO}&=\frac{m\als ^5 C_F^4 (8 T_Fn_l-C_A) }{432 \pi  n^3}\nn\\
&+\frac{m\als^6 C_F^4  }{1944 \pi ^2n^3}\left[27 C_F T_Fn_l-150 C_A T_Fn_l-\frac{3}{8}  \left(24 \pi ^2-59\right) C_A^2\right.\nn\\
&\left.-\frac{9}{4} (8 T_Fn_l-C_A) \left(-2 C_F+2 C_A \left(\ln \frac{n }{ \alpha  C_F }-H_{n+1}+\frac{3}{2 n}+1\right)\right.\right.\nn\\
&\left.\left.+\beta_0 \left(5\ln \frac{\mu  n}{ \alpha  C_F m}-5H_{n-2}+4 H_{n+1}-2 n \psi ^{(1)}(n-1)-\frac{7}{n}+\frac{23}{3}+\frac{ n^4+7}{n \left(n^2-1\right)^2}\right)\right)\right.\nn\\
&\left.+\pi ^2 C_F^2  \left(-\frac{81}{5n^2}+\frac{162}{5n}+\frac{1233}{25}\right)\right]
\,.
\end{align}

\subsection{Matching-scheme dependence}
\label{Sec:offon}

It is possible to compute $\Delta$ with N$^4$LO accuracy for the equal mass case using the potential obtained in Ref.\ \cite{Mishima:2024afk} in the on-shell matching scheme. Their result for $\tilde V_4$ reads:
\begin{align}
\tilde V_{4,\text{on-shell}}^{S^2,\text{P}}&=\frac{C_F\pi\als}{m^4}\frac{{\bf S}_1\cdot(\vk\times \vp){\bf S}_2\cdot(\vk\times \vp)}{k^2}-\frac{2(\pi  \als) ^2 C_F^2}{3 m^3} k {\bf S}_1\cdot{\bf S}_2\nn\\
&-\frac{2\als ^3 C_F {\bf S}_1\cdot{\bf S}_2 }{3 \pi  m^2}\left[
\left(\frac{7 C_A}{2}-\beta_0\right)\left(\ln^2 k(\beta_0-C_A)- 2\ln k(\beta_0\ln \mu-C_A \ln m) \right)\right.\nn\\
&\left.+\frac{1}{3} \ln k \left(2 \left(11+\pi ^2\right) C_A^2-\frac{1}{2} (3 C_A) (16 \beta_0+25 C_F)+\frac{1}{2} \beta_0 (10 \beta_0+21 C_F)\right)\right]
\,.
\label{deftV4HFPOS}
\end{align}
The two schemes are related by the following equality:
\begin{align}
\tilde{V}_4^{S^2,\text{P}}=\tilde V_{4,\text{on-shell}}^{S^2,\text{P}}+\delta\tilde{V}_4^{S^2,\text{P}}
\,,
\end{align}
where all the NRQCD Wilson coefficients in \eqref{deftV4HFP} have been evaluated at LO,
with
\begin{align}
\delta\tilde{V}_4^{S^2,\text{P}}&=\tilde V_{S^2,2}^{(4)}+\tilde V_{S^2,3}^{(4)}+\tilde V_{S^2,5}^{(4)}-\frac{1}{2}\tilde V_{S^2}^{(3)}.
\end{align}
Using the relation in \eqref{relkVoff}, it is straightforward to see that, as expected, $\langle 
\delta\tilde{V}_4^{S^2,\text{P}}\rangle_\text{UF}=0$, which proves the scheme-independence of the result. 

\subsection{Phenomenology}

We now study the impact of our findings, building on the phenomenological analysis presented in Ref.~\cite{Peset:2018jkf}. To the N$^3$LO and N$^3$LL results discussed there, we add new lines with the complete N$^4$LO result and with the partial N$^4$LL result. For the partial N$^4$LL result, we include the running of all NRQCD Wilson coefficients except for $c_{W_2}$, which we set to zero. This choice is due to the gauge-dependence of $c_{W_2}$, which should instead be always be considered in conjunction with the Wilson coefficient of the appropriate four-fermion operator. 

\begin{figure}[H]
\begin{center}
\includegraphics[width=.8\textwidth]{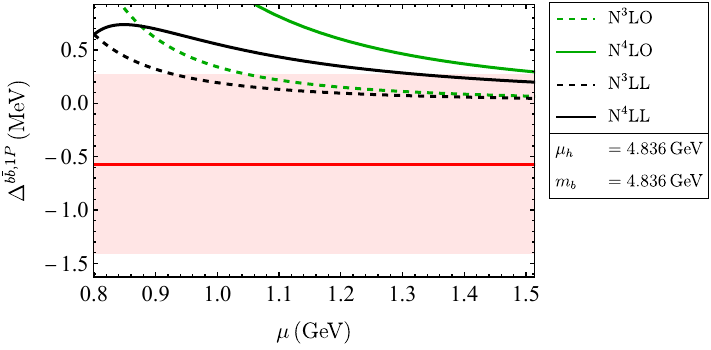}
\end{center}
\caption{P-wave hyperfine splitting for $n=2$ bottomonium. Experimental line and error band in red. Theory predictions are carried out with the PV mass $m_{b,\rm PV}=4.836$ GeV quoted from \cite{Ayala:2019hkn}.  }
\label{Fig:b2}
\end{figure}

\begin{figure}[H]
\begin{center}
\includegraphics[width=.8\textwidth]{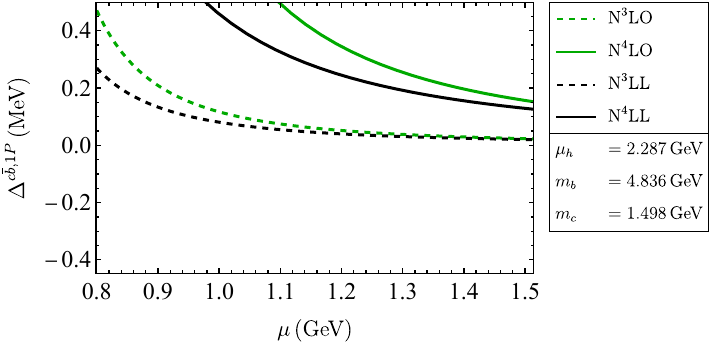}
\end{center}
\caption{P-wave hyperfine splitting for $n=2$ $B_c$. Theory predictions are carried out with the PV mass $m_{b,\rm PV}=4.836$ GeV quoted from \cite{Ayala:2019hkn} and $m_{c,\rm PV}
=1.498$ GeV. The latter is obtained using the experimental mass of the $D$ mesons and the value of $\bar \Lambda$ determined in \cite{Ayala:2019hkn}. }
\label{Fig:bc}
\end{figure}

\begin{figure}[H]
\begin{center}
\includegraphics[width=.85\textwidth]{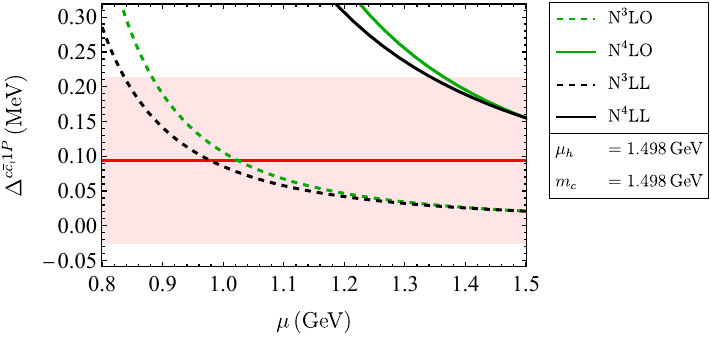}
\end{center}
\caption{P-wave hyperfine splitting for $n=2$ charmonium. Experimental line and error band in red. Theory predictions are carried out with the PV mass $m_{c,\rm PV}=1.498$ GeV, obtained using the experimental mass of the $D$ mesons and the value of $\bar \Lambda$ determined in \cite{Ayala:2019hkn}. }
\label{Fig:c2}
\end{figure}

\begin{figure}[H]
\begin{center}
\includegraphics[width=.8\textwidth]{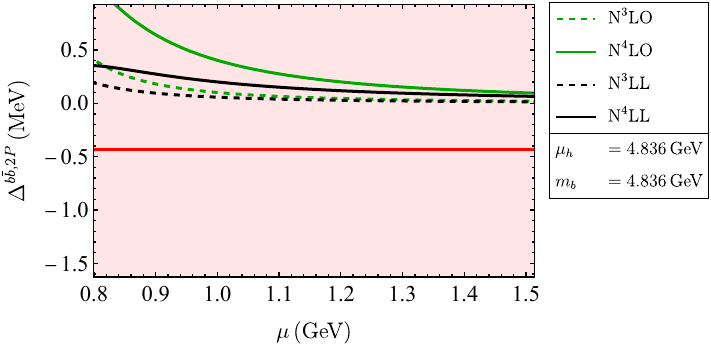}
\end{center}
\caption{P-wave hyperfine splitting for $n=3$ bottomonium. Experimental line and error band in red. Theory predictions are carried out with the PV mass $m_{b,\rm PV}=4.836$ GeV quoted from \cite{Ayala:2019hkn}. 
}
\label{Fig:b3}
\end{figure}

Our results are shown in Figs.~\ref{Fig:b2}, \ref{Fig:bc}, and \ref{Fig:c2} for $n=2$ P-wave heavy quarkonium states in bottomonium, $B_c$, and charmonium. We also consider $n=3$ P-wave bottomonium states, with the results displayed in Fig.~\ref{Fig:b3}. Our main target is the $n=2$ P-wave bottomonium, where we expect the weak-coupling computation to be most reliable. Nevertheless, we also explore the other states to evaluate the performance of a weak-coupling analysis more broadly. 

It is worth emphasizing that relativistic corrections to this observable are incorporated for the first time and are expected to be sizable.  This brief phenomenological analysis confirms this expectation. The N$^4$LO correction is larger than the leading non-vanishing order, suggesting that the N$^3$LO result may have been anomalously small due to an accidental cancellation between Abelian and non-Abelian contributions and that the natural size of the effect is likely given by the N$^4$LO result. For bottomonium, the absolute magnitude of this correction remains small, but it is significantly larger for charmonium and $B_c$. 

On the other hand, the fixed N$^4$LO result exhibits a large renormalization scale dependence. Remarkably, the partial resummation of logarithms substantially improves this scale dependence for bottomonium, though it has a smaller effect for $B_c$ and charmonium. For bottomonium, the agreement with experiment is reasonable, whereas for charmonium, the results worsen when moving to smaller renormalization scales, which are more natural for the bound state. Before drawing any definitive conclusion the complete N$^4$LL result is required.

In all the above mentioned plots, $\alpha_s$ runs with $n_l=3$ active flavors as provided by the RunDec code \cite{Chetyrkin:2000yt}, with $\alpha_s^{(n_l = 5)}(91.1876 \text{ GeV})=0.1184$ (with decoupling
at $\overline{m}_b = 4.2$ GeV and at $\overline{m}_c = 1.27$ GeV; the specific location of the decoupling plays a
marginal role in the plots). The resummation of the hard logarithms is performed at the hard-scale choice $\mu_h=2m_r$ for each system.

\section{QED}
\label{sec:QED}
By changing the colour factors, we can easily apply the results above to QED bound states such as positronium, muonium, dimuonium and muonic hydrogen. For positronium/dimuonium, they yield
\begin{align}
\Delta_\text{pos}&=\frac{m\alpha ^6  }{1440 n^5}\left[c_F^4 \left(\frac{287 n^2}{60}+\frac{3 n}{2}-\frac{5}{2}\right)+c_F^2 \left(\frac{335 n^2}{9}+10 n-20\right)+2 c_Fc_{pp'}\left(2 n^2-3\right)\right.\nn\\
&\left.+c_Fc_S \left(-\frac{97 n^2}{9}+10 n+12\right)-40 c_Fc_{W_2}\left(n^2-1\right)+c_S^2 \left(\frac{191 n^2}{36}+\frac{5 n}{2}-\frac{3}{2}\right)\right]
\,.
    \label{Deltapos}
\end{align}

For muonium or hydrogen (assuming a point-like proton), we obtain
\begin{align}
\Delta_\text{H}&=\frac{m_r^5\alpha ^6 }{45 m_1^2 m_2^2 n^5}\left[c_F^{(1)2} c_F^{(2)2} \left(\frac{287 n^2}{60}+\frac{3 n}{2}-\frac{5}{2}\right)
\right.\nn\\
&
+\left(2 n^2-3\right) \left(\frac{c_F^{(1)} c_{pp'}^\two m_1}{m_2}+\frac{c_F^{(2)} c_{pp'}^\one m_2}{m_1}-\frac{c_S^{(1)} c_S^{(2)}}{2}\right)
\nn\\
&\left.-2 \left(3 n^2-2\right) \left(-2 c_F^{(1)} c_F^{(2)}+c_F^{(1)} c_S^{(2)} \left(\frac{m_1}{m_r}+1\right)+c_F^{(2)} c_S^{(1)} \left(\frac{m_2}{m_r}+1\right)\right)\right.\nn\\
&\left.+\left(\frac{227 n^2}{36}+\frac{5 n}{2}-3\right) \left(c_F^{(1)}+c_F^{(2)}+\frac{c_S^{(1)} m_2}{2 m_1}+\frac{c_S^{(2)} m_1}{2 m_2}\right)^2\right.\nn\\
&\left.-20 \left(n^2-1\right) \left(\frac{c_{W_2}^\two c_F^{(1)} m_1}{m_2}+\frac{c_{W_2}^\one c_F^{(2)} m_2}{m_1}\right)\right]
\,.
    \label{Deltamuo}
\end{align}
For muonic hydrogen, assuming a point-like proton and that the electron mass is smaller than the inverse Bohr radius, we obtain 
\begin{align}
\Delta_{\mu\text{H}}&=\frac{4 \alpha ^5 c_F^\one c_F^\two m_r^3}{27 \pi  m_1 m_2 n^3}\nn\\
&+\frac{8 \alpha ^6 m_r^3 }{81 \pi ^2 m_1 m_2 n^3}\left[-c_F^\one c_F^\two \left(5 \ln \frac{m_e n}{2 \alpha  m_r}-5 H_{n-2}+4 H_{n+1}-2 n \psi ^{(1)}(n-1)\right.\right.\nn\\
&\left.\left.+\frac{n^4+7}{n \left(n^2-1\right)^2}-\frac{7}{n}+\frac{157}{24}\right)+\frac{9 \pi ^2 m_r^2 }{40 m_1 m_2 n^2}\left(c_F^{(1)\,2} c_F^{(2)\,2} \left(\frac{287 n^2}{60}+\frac{3 n}{2}-\frac{5}{2}\right)\right.\right.\nn\\
&\left.\left.+\left(2 n^2-3\right) \left(\frac{c_F^\one c_{pp'}^\two m_1}{m_2}+\frac{c_F^\two c_{pp'}^\one m_2}{m_1}-\frac{c_S^\one c_S^\two}{2}\right)\right.\right.\nn\\
&\left.\left.+2 \left(3 n^2-2\right) \left(2 c_F^\one c_F^\two-c_F^\one c_S^\two \left(\frac{m_1}{m_r}+1\right)+c_F^\two (-c_S^\one) \left(\frac{m_2}{m_r}+1\right)\right)\right.\right.\nn\\
&\left.\left.+\left(\frac{227 n^2}{36}+\frac{5 n}{2}-3\right) \left(c_F^\one+c_F^\two+\frac{c_S^\one m_2}{2 m_1}+\frac{c_S^\two m_1}{2 m_2}\right)^2\right.\right.\nn\\
&\left.\left.-10 \left(n^2-1\right) \left(\frac{2 c_{W_2}^\two c_F^\one m_1}{m_2}+\frac{2 c_{W_2}^\one c_F^\two m_2}{m_1}\right)\right)\right],
    \label{DeltamuH}
\end{align}
where in the last equation the running $\alpha$ has been transformed into the physical electromagnetic $\alpha$ and $m_e$ is the electron mass.

At N$^4$LO we find:
\begin{align}
\label{th:pos}
  \Delta_\text{pos}^\text{N$^4$LO}&=  \frac{m\alpha ^6  \left(137 n^2+90 n-45\right)}{5400 n^5},
\end{align}
\begin{align}
\nn
   \Delta_\text{H}^\text{N$^4$LO}&=  \frac{m_r^5\alpha ^6 }{5400 m_1^2 m_2^2 n^5}\left[2419 n^2+1530 n-780+\frac{5}{6} \left(227 n^2+90 n-108\right) \left(\frac{m_1^2}{m_2^2}+\frac{m_2^2}{m_1^2}\right)\right.\\
   &\left.+\frac{20}{3} \left(119 n^2+90 n-36\right) \left(\frac{m_1}{m_2}+\frac{m_2}{m_1}\right)\right],
\end{align}
and
\begin{align}
  \Delta_{\mu\text{H}}^\text{N$^4$LO}&=\frac{4 \alpha ^5 m_r^3}{27 \pi  m_1 m_2 n^3}-\frac{8 \alpha ^6 m_r^3 }{81 \pi ^2 m_1 m_2 n^3}\left[5 \ln \frac{m_e n}{2 \alpha  m_r}-5 H_{n-2}+4 H_{n+1}-2 n \psi ^{(1)}(n-1)\right.\nn\\
  &\left.+\frac{n^4+7}{n \left(n^2-1\right)^2}-\frac{7}{n}+\frac{121}{24}-\frac{9 \pi ^2}{20}  \left(\frac{1}{2 n^2}+\frac{5}{4 n}+\frac{11}{72}\right)\right.\nn\\
  &\left.-\frac{9 \pi ^2 m_r^2 }{40 m_1 m_2 n^2}\left(\frac{454 n^2}{45}+4 n-4+\left(\frac{m_1 m_2}{2 m_r^2}\right)^2\left(\frac{227 n^2}{36}+\frac{5 n}{2}-3\right) \right)\right].
\end{align}
Numerically, they yield 
\begin{align}
   \Delta_{\text{pos}}= 0.46 \text{ MHz}, \quad
   \Delta_{\text{dimuon}}= 95.81\text{ MHz},\quad
   \Delta_{\text{muon}}=0.55\text{ MHz} ,
\end{align}
for positronium, dimuonium and muonium respectively.

The above results provide highly nontrivial tests of intricate QED computations performed over the years for positronium.
We find that our result agrees with the corresponding expression in Refs.~\cite{Khriplovich:1993zz,Czarnecki:1999mw,Zatorski} but disagrees with the most recent computation in Ref.\ \cite{Patkos:2024lqf}. After communicating our results to the authors of Ref.~\cite{Patkos:2024lqf}, a mistake was found in their computation, and they subsequently obtain agreement with the older results.

Our computation also readily accommodates the case of unequal masses, which can be of interest for muonium and hydrogen systems. Furthermore, our computations can also be of interest for the case of muonic hydrogen, as they incorporate the effect of a light electron, allowing for straightforward identification of logarithmic contributions. Nevertheless, to make our results fully applicable, the effects associated with the finite electron mass should also be incorporated. 



\section{Conclusions}

In this paper, we have computed the $1/m^2$ spin-dependent and velocity-independent potential to two-loops in the Wilson-loop matching scheme with general $D$ dimensions, and discussed its relation to computations in other matching schemes. We have also computed the $1/m^3$ potential to one-loop for general $D$ dimensions within the off-shell matching scheme, correcting earlier results. In addition, we have provided the $1/m^4$ tree-level potential relevant for spin-dependent observables and discussed the relations among different matching schemes. These results constitute the necessary ingredients for future higher-order computations of spin-dependent corrections to the spectrum. 

Using these ingredients, we have computed the ultrafine splitting with N$^4$LO accuracy for heavy quarkonium, as well as for analogous systems in atomic physics. Our results are also relevant for achieving N$^4$LL precision, which is now within reach. As a first taste of its importance, we have presented a partial N$^4$LL result that incorporates the hard logarithms associated with the Wilson coefficients of the bilinear terms in the NRQCD Lagrangian. Notably, with N$^4$LO/N$^4$LL precision, relativistic corrections to this observable are incorporated for the first time, and they are expected to be sizable. A brief phenomenological analysis supports this expectation. 

The fixed-order result exhibits a strong renormalization-scale dependence, as expected from its proportionality to a high power of $\als$. Remarkably, the partial resummation of logarithms implemented in this work significantly neutralizes this scale dependence in the case of bottomonium (though not for charmonium or $B_c$). A comprehensive phenomenological analysis is deferred until the full N$^4$LL evaluation of the ultrafine splitting becomes available.

We have also applied our results to analogous QED observables relevant to atomic physics. In this context, our findings have shed light on a recent conflict between two different results reported in the literature for the spectrum of P-wave states.

\medskip
   
\noindent
{\bf Acknowledgments.}\\
We thank conversations with K. Pachucki, V. Patkos, A. Penin, M. Steinhauser, Y. Sumino and J. Zatorski. This work was supported in part by the Spanish Ministry of Science and Innovation Grant No. PID2023-146142NB-I00 and PID2022-136510NB-C31 funded by MCIN/AEI/ 10.13039/501100011033.


\appendix

\section{Treatment of the Dirac and Pauli matrices in $D$ dimensions}
\label{App:Spin}

It is not straightforward how to treat spin, that is, the Dirac and Pauli matrices, in $D$ dimensions. The following equalities hold independently of the specific treatment of the Pauli matrices:
\begin{align}
    \{\sigma^i,\sigma^j\}=2\delta^{ij},\qquad \delta^{ii}=D-1=d.
\end{align}

Nevertheless, the treatment of the Levi-Civita tensor in $D$ dimensions is not unique. This becomes relevant once divergent loops appear. One must be specially careful to maintain consistent conventions across the different contributions to heavy quarkonium observables, as they originate from distinct energy scales. 

At the hard scale, the convention used in Ref.~\cite{Pineda:1998kj} (see eq.~(5.10)) defines
\begin{align}
[\sigma^i, \sigma^j] = 2 i \epsilon^{ijk} \sigma^k, \qquad \epsilon^{ijk} \epsilon^{ijl} = (D-2) \delta^{kl}.
\end{align}

A variant of this convention is to work with three-dimensional Pauli matrices, modifying the last equality of eq.~(5.10) in Ref.~\cite{Pineda:1998kj} by setting $D-2 \to 2$ (see the discussion in Ref.~\cite{Pineda:2000sz}). In this variant, one uses
\begin{align}
[\sigma^i, \sigma^j] = 2 i \epsilon^{ijk} \sigma^k, \qquad \epsilon^{ijk} \epsilon^{ijl} = 2 \delta^{kl}.
\end{align}
 
 Another possibility is to project to the spin one and spin zero states, following the approach of Ref.~\cite{Czarnecki:1999mw}. This method was also employed in the N$^3$LL evaluation of the hyperfine splitting of heavy quarkonium \cite{Kniehl:2003ap,Penin:2004xi} (which includes the two-loop soft running of the spin-dependent delta potential). It was also used in the partial computation of the two-loop soft running of the spin-independent delta potential in Ref.\ \cite{Anzai:2018eua}, where it was found that, at one-loop, this method reproduces the same result as in Ref.\ \cite{Peset:2015vvi}, which uses the convention of Ref.~\cite{Pineda:1998kj}. 

This last method is the one we adopt in this paper. It relies on projectors to isolate spin-singlet and spin-triplet states, avoiding any need to define the commutator of Pauli matrices or use the Levi-Civita tensor in $D$ dimensions. A key point in this discussion is at which point the cancellation of divergences takes place. Here, it can be performed before the bound state dynamics takes place (as it can be factored out). Consequently, the bound-state computation itself can be carried out in four dimensions after divergences have canceled.

This discussion also applies to the spin, where they can be expressed in terms of $s$, the particle spin, treated as in four dimensions, multiplied by the potentials defined in $D$ dimensions. However, this is strictly valid only when the potential appears at leading order in the computation. If multiple iterations of spin-dependent potentials are considered, projection onto spin states can only be performed at the final stage. Note also that handling non-diagonal potentials in spin-space (like $S_{12}$) is more complicated with this method. 

The spin projectors have been defined in Ref.~\cite{Czarnecki:1999mw} in a four-dimensional representation. For singlet and triplet states they read respectively\footnote{Note that in the original publication $P_S\equiv\Psi_P$ and $P_T=\Psi_O$.}
\begin{align}
    P_S=\frac{1+\gamma^0}{2\sqrt{2}}\gamma^5,\quad P_T=\frac{1+\gamma^0}{2\sqrt{2}}\bfgamma\cdot {\hat n}
    \,,
\end{align}
where ${\hat n}$ stands for the direction of the helicity. Then the contribution of an operator consisting of a chain of Dirac matrices $\Gamma$ to a singlet or triplet state is given by
\begin{align}
    &{\rm Tr}(P_S^\dagger \Gamma P_S\Gamma)=\frac{1}{2}{\rm Tr}(\Gamma_{11}\Gamma_{22}),\nn\\
& {\rm Tr}(P_T^\dagger \Gamma P_T\Gamma)|_{\xi-{average}}=\frac{1}{D-1}{\rm Tr}({\bf P_T}^\dagger \Gamma {\bf P_T}\Gamma)=\frac{1}{2(D-1)}{\rm Tr}(\sigma^i\Gamma_{11}\sigma^i\Gamma_{22})
\,,
\end{align}
where ${\bf P_T}=\frac{1+\gamma^0}{2\sqrt{2}}{\bfgamma}$ and, in the last equality of each equation, we have expressed it in terms of two-dimensional Pauli matrices given by
\begin{align}
  \Gamma=\begin{pmatrix}
\Gamma_{11}&\Gamma_{12}\\
\Gamma_{21}&\Gamma_{22}
\end{pmatrix}.  
\end{align}

The expression in terms of Pauli matrices agrees with eqs.~(4.90)-(4.91) of \cite{Beneke:2013jia}.

We now want to show how operators such as 
\begin{align}
    \mathcal{O}_0&=\Sigma_0^{(1)}\otimes \Sigma_0^{(2)}=1\otimes 1\nn\\
\mathcal{O}_1&=\Sigma_1^{(1)}\otimes \Sigma_1^{(2)}=-\frac{1}{8}[\sigma^i,\sigma^j]\otimes [\sigma^i,\sigma^j]\nn\\
\mathcal{O}_2&=\Sigma_2^{(1)}\otimes \Sigma_2^{(2)}=\frac{1}{64}[\sigma^i,\sigma^j][\sigma^k,\sigma^l]\otimes [\sigma^i,\sigma^j][\sigma^k,\sigma^l].
\end{align}
are expressed in this convention. These operators appear in \cite{Gerlach:2019kfo} when computing the four-fermion hard matching coefficients. We will not explicitly dwell with the hard computation in this paper (though such expressions will be needed in forthcoming papers) but similar expressions appear in the soft computation of the potentials we carry out in this paper. 

With the above considerations, we find
\begin{align}
\mathcal{O}_0&=1\otimes 1=1,\nn\\
\mathcal{O}_1&=\left\{\begin{matrix}
\frac{(D-2)(D-1)}{2}&\text{ for singlet}\\
\frac{(D-2)(D-5)}{2}&\text{ for triplet}
\end{matrix} \right.=\frac{(D-2)(D-4)}{2}-2(D-2){\bf S}_1\cdot{\bf S}_2\nn\\
\mathcal{O}_2&=\left\{\begin{matrix}
\frac{(D-2)(D-1)}{4} \left(D^2-11 D+26\right)&\text{ for singlet}\\
\frac{(D-2)(D-5)}{4} \left(D^2-15 D+34\right)&\text{ for triplet}
\end{matrix} \right.\nn\\
&=\frac{D-2}{4}\left(D^3-18 D^2+91 D-134\right)-2 (D-6) (D-3) (D-2){\bf S}_1\cdot{\bf S}_2
\,,
\end{align}
where ${\bf S}_1\cdot{\bf S}_2=-3/4$ for spin-singlet and ${\bf S}_1\cdot{\bf S}_2=1/4$ for spin-triplet states. We have also checked these expressions using FeynCalc \cite{FeynCalc4,FeynCalc3}.

\section{NRQCD Feynman rules}
\label{App:FR}

We give here the NRQCD Feynman rules used for the computation of $V_{S^2}$ to two loops in Sec. \ref{Sec:WL} and the $1/m^3$ potential to one loop in Sec. \ref{Sec:m3}. Several of these can already be found in Ref.~\cite{Pineda:2011dg}. We rewrite these in a more suitable way for computations in $D$ dimensions. The NRQCD vertex proportional to $\sim \frac{c_S}{m^2}\partial_0{\bf A}$ has been eliminated by using the equations of motion. This changes some of the Feynman rules below and generate extra terms at ${\cal O}(1/m^3)$, which we do not explicitly display here.

\subsection{Preliminaries}

\begin{itemize}
\item All momenta in vertices are
  \emph{incoming}. The only propagators where the direction matters
  are the fermionic ones, where the momentum is taken in the direction
  of the arrow.
\item The covariant derivative is
  defined as $D = \partial + i g A$. 
\item For the heavy \emph{antiquark} we use a charge-conjugated field,
  see Ref.~\cite{Pineda:2011dg}.
\item All indices are strictly Euclidean, i.e.\ ${\bf p}\cdot {\bf q} = p^i  q^i$.
\item The gluon propagator is computed in the Feynman gauge.
\end{itemize}

\subsection{Propagators}

\begin{tblr}{
    colspec = {cl},
    colsep = 1em,
    rowsep = 1em,
  }
  \raisebox{-9.4pt}{\includegraphics{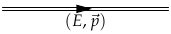}} & $i \frac{1}{E + m - \sqrt{\vec{p}^2 + m^2} + i\eta}\simeq \frac{i}{E+i\eta}$\\
  \raisebox{-9.4pt}{\includegraphics{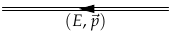}} & $i \frac{1}{E + m - \sqrt{\vec{p}^2 + m^2} + i\eta}\simeq \frac{i}{E+i\eta}$\\
  \raisebox{-6.8pt}{\includegraphics{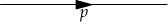}} & $i \frac{\slashed{p}}{p^2 + i\eta}$\\
  \raisebox{-11.3pt}{\includegraphics{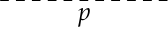}} & $-i \frac{1}{p^2 + i\eta}$\\
  \raisebox{-11.3pt}{\includegraphics{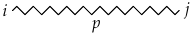}} & $i \frac{\delta^{ij}}{p^2 + i\eta}$\\
  \raisebox{-11.3pt}{\includegraphics{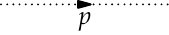}} & $i \frac{1}{p^2 + i\eta}$\\
\end{tblr}

The last one is the ghost propagator, dashed represents the zero component of the gluon field and zigzag line the space-components of the gluon field. 

\subsection{Vertices}

\DefTblrTemplate{middlehead,lasthead}{default}{}
\DefTblrTemplate{firstfoot,middlefoot}{default}{}
\begin{longtblr}[
  label = none,
  entry = none,
  caption = {},
  ]{
    colspec = {Q[c,c]Q[c,l]},
    colsep = 1em,
    rowsep = 1em,
  }
  \includegraphics{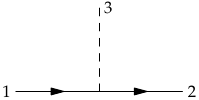} & $-i g T^{a_3}_{j_2 j_1} \gamma^0$ \\
  \includegraphics{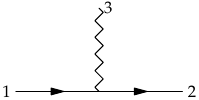} & $ig T^{a_3}_{j_2 j_1} \gamma^{i_3}$\\
 \includegraphics{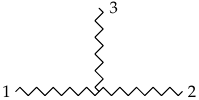} & $
    -g f^{a_1a_2a_3}\big[\delta^{i_1i_2}(p_1-p_2)^{i_3}
    + \delta^{i_2i_3}(p_2-p_3)^{i_1} + \delta^{i_3i_1}(p_3-p_1)^{i_2}\big]
  $ \\
  \includegraphics{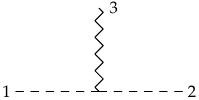} & $g f^{a_1a_2a_3} (p_1 - p_2)^{i_3}$\\
  \includegraphics{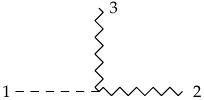} & $g f^{a_1a_2a_3} \delta^{i_2i_3}(p_2-p_3)_{0}$\\
  \includegraphics{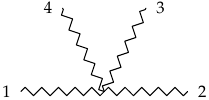} & {$-i g^2[V(1,2,3,4) + V(1,3,2,4) + V(1,4,2,3)]$\\[1em]
  $V(1,2,3,4) = f^{a_1a_2c}f^{a_3a_4c}(\delta^{i_1i_3}\delta^{i_2i_4} - \delta^{i_1i_4}\delta^{i_2i_3})$}\\
  \includegraphics{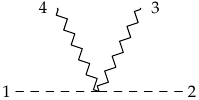} & $ig^2\delta^{i_3i_4} (f^{a_1a_3c}f^{a_2a_4c} + f^{a_1a_4c}f^{a_2a_3c})$ \\
  \includegraphics{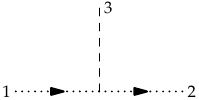} & $-g f^{a_1a_2a_3} (p_2)_0$ \\
  \includegraphics{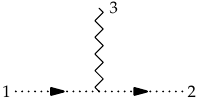} & $g f^{a_1a_2a_3} (p_2)^{i_3}$\\
    \includegraphics{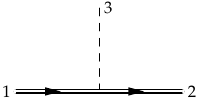} & $-i g T^{a_3}_{j_2 j_1} \left(1 - \frac{1}{8m^2} \left(c_D {\bf p}_3^{\,2} - c_S [\bfsigma \cdot {\bf p}_1, {\bfsigma} \cdot {\bf p}_2]\right)\right)$\\
    \includegraphics{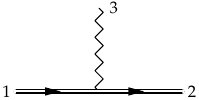} & $i \frac{g}{2m} T^{a_3}_{j_2 j_1} \left((p_1 - p_2)^{i_3} - \frac{c_F}{2} [\sigma^{i_3}, \bfsigma \cdot {\bf p}_3]\right)$
    \\
    \includegraphics{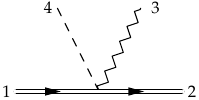} & $
   -i \frac{g^2}{4m^2} \left(
     c_D (p_4)^{i_3} [T^{a_3}, T^{a_4}]_{j_2 j_1}
      + \frac{c_S}{2} [\sigma^{i_3}, \bfsigma \cdot {\bf p}_4]\{T^{a_3}, T^{a_4}\}_{j_2 j_1}
      \right)$
     \\ 
    \includegraphics{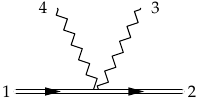} &
    $\begin{aligned}
    &
      -i\frac{g^2}{2m} \Big[
      \delta^{i_3i_4} \{T^{a_3}, T^{a_4}\}_{j_2 j_1}
      + \frac{c_F}{2} [\sigma^{i_3}, \sigma^{i_4}] [T^{a_3}, T^{a_4}]_{j_2 j_1}
      \\&
      \quad
      + \frac{(p_3 - p_4)_0}{4m} \Big(
      c_D \delta^{i_3i_4}[T^{a_3}, T^{a_4}]_{j_2 j_1}
      + \frac{c_S}{2} [\sigma^{i_3}, \sigma^{i_4}]\{T^{a_3}, T^{a_4}\}_{j_2 j_1}
      \Big)
      \Big]
    \end{aligned}
    $
      \\
    \includegraphics{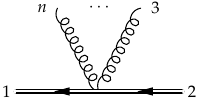} &  $= \; (-1)^n \times \left.\includegraphics{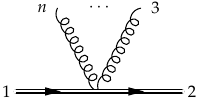}\right \rvert_{T^a \to (T^a)^T}$
    $\left(\begin{minipage}{110pt}
Any combination of scalar and vector gluons. Sign corresponds to $g \to -g$
    \end{minipage}\right)$
   \\
\end{longtblr}

Note that the last Feynman rule applies for the physical antiparticles, and for the time or space component of the four-vector gluon. To write the Pauli matrices in the way that appear in Appendix~\ref{App:Spin}, one has to apply charge conjugation to the antiparticles spin. The net effect of this is that the spin matrices get transposed.

\section{Expectation values}
\label{App:ME}

To obtain \cref{EnHFPatN4LO} we need the following set of expectation values:

\subsection{Double insertions}
The expectation values of double potential insertions are better computed with position space potentials. For the reduced Green's function we use the definition provided in ref.~\cite{Kiyo:2014uca}.
\begin{align}
\langle nP| \frac{1}{r}\overline{G}_n\text{reg}\frac{1}{r^3}|nP\rangle&=\frac{ m_r}{ 2 a^2n^3},\\
\langle nP| \frac{\ln(re^{\gamma_E})}{r}\overline{G}_n\text{reg}\frac{1}{r^3}|nP\rangle&=\frac{ m_r}{6 a^2 n^3}\left(3 \ln\frac{a n}{2}+H_{n-2}-2 n \psi ^{(1)}(n-1)\right.\nn\\
&\left.\hspace{2cm}+\frac{4 n^5+27 n^4-8 n^3-12 n^2+4 n+9}{3 n \left(n^2-1\right)^2}\right)\\
\langle nP|\frac{1}{r^3}\bar{G_n}\frac{1}{r^3}|nP\rangle&=\frac{m_r}{a^4n^3}\left(-\frac{1}{5  n^2}+\frac{1}{6  n}+\frac{227}{540 }\right),
\end{align}
where $a=1/(m_r\als C_F)$.

\subsection{Single insertions}

To evaluate the expectation values of \cref{deftV3HFP,deftV4HFP}, we require the expectation values derived in this section. These are computed most conveniently in momentum space. Obviously, choosing position or momentum space representation for the potential does not change the final result.

\subsubsection{Tree level $1/m^4$ and $1/m^3$ potentials}

After using \eqref{HFspinproj}, we need only the expectation values:
\begin{align}
    \langle nP|\vp^2|nP\rangle&=\langle nP|\vp'^2|nP\rangle=0,\\
    \langle nP|\vk^2|nP\rangle&=\frac{2 \left(1-n^2 \right)}{3 \pi  a^5 n^5},\\
   \langle nP| \frac{(\vp^2-\vp'^2)^2}{\vk^2}|nP\rangle&=-\frac{\pi}{a} \langle nP|k|nP\rangle,\label{relkVoff}\\
    \langle nP|k|nP\rangle&=\frac{2 \left(2-3 n^2\right)}{15 \pi ^2 a^4 n^5},
\end{align}
where the third relation is obtained from \cite{Adkins:2019yix}.

\subsubsection{Two loop $1/m^2$ potential} 
For an $nP$ state we have the following equalities:
\begin{align}
\langle nP|\ln k|nP\rangle&=-\frac{1}{4\pi}\langle nP|\text{reg}\frac{1}{r^3}|nP\rangle=-\frac{1}{12 \pi  a^3 n^3},\\
\langle nP|\ln^2 k|nP\rangle&=\frac{1}{2\pi}\langle nP|\text{reg}\frac{\ln (re^{\gamma_E})}{r^3}-\text{reg}\frac{1}{r^3}|nP\rangle\nn\\
&=\frac{1}{6 \pi  a^3 n^3}\left(\ln\frac{a n}{2}-H_{n+1}-\frac{2 n-\frac{3}{2}}{n}+\frac{37}{12}\right).
\end{align}

\bibliographystyle{JHEP} 
\bibliography{Bibliography} 
\end{document}